\documentclass{article}

\usepackage[nonatbib,preprint]{neurips_2019}
\usepackage[numbers]{natbib}

\usepackage[utf8]{inputenc} 
\usepackage[T1]{fontenc}    
\usepackage{hyperref}       
\usepackage{url}            
\usepackage{booktabs}       
\usepackage{amsfonts}       
\usepackage{nicefrac}       
\usepackage{microtype}      

\usepackage{graphicx,calc}
\newlength\myheight
\newlength\mydepth
\settototalheight\myheight{Xygp}
\newcommand*\inlinegraphics[1]{%
	\settototalheight\myheight{Xygp}%
	\settodepth\mydepth{Xygp}%
	\raisebox{-\mydepth}{\includegraphics[height=\myheight]{#1}}%
}
\usepackage{amsmath} 
\newtheorem{theorem}{Theorem}
\newtheorem{Proposition}{Proposition}

\usepackage{subcaption} 
\usepackage{xcolor}
\usepackage{wrapfig}

\title{Measuring the Clustering Strength of a Network via the Normalized Clustering Coefficient}

\author{
	Ting Li\\
	HKUST\\
	\texttt{tlial@ust.hk} \\
	\And
	Xianshi Yu \\
	University of Michigan \\
	\texttt{xsyu@umich.edu} \\
	\And
	Bing-Yi Jing \\
	HKUST\\
	\texttt{majing@ust.hk} \\
}

\begin{document}
	
	\maketitle
	
	\begin{abstract} 
In this paper, we propose a novel statistic of networks, the normalized clustering coefficient, which is a modified version of the clustering coefficient that is robust to network size, network density and degree heterogeneity under different network generative models. In particular, under the degree corrected block model (DCBM), the "in-out-ratio" could be inferred from the normalized clustering coefficient. Asymptotic properties of the proposed indicator are studied under three popular network generative models. The normalized clustering coefficient can also be used for networks clustering, network sampling as well as dynamic network analysis. Simulations and real data analysis are carried out to demonstrate these applications.
	\end{abstract}
	
	\section{Introduction}
	Nowadays, complex networks are ubiquitous in the real world. 
Network summary statistics have been widely used in experimental and empirical studies. Some of these are model-free such as network size, density, average shortest path and so on. Some are related to generative models' parameters, for instance, the number of blocks for the stochastic block model (SBM)(\cite{sbm}).
	
Many real-world networks show properties of small-world networks (\cite{watts1998}) as their neighborhood connectivity is higher than that in comparable random networks. To measure the clustering strength, the clustering coefficient is proposed in graph theory. Two versions of the measure exist: the global one and the local one. The global clustering coefficient was proposed in \cite{luce1949} and is
	\begin{equation}\label{cc}
	\hat{cc}=\frac{3\times \mbox{number of triangles}}{\mbox{number of all triplets (open and closed)}}.
	\end{equation}
	It was extended to weighted networks by \cite{opsahl2009}. The local clustering coefficient was introduced by \cite{watts1998} and generalized to weighted, directed, and bipartite graphs (\cite{barrat2004}, \cite{latapy2008}, \cite{clemente2018}).
	
	However, the clustering coefficient is not without problems. For example, under the degree corrected block model (DCBM), $\hat{cc}$ not only tends to $0$ when the network is large and sparse, but also involves noise from degree heterogeneity (more details in Section \ref{tNCC} below). To overcome these issues, we propose a so-called normalized clustering coefficient:
	$$\hat{\rho}=\frac{\hat{T}\hat{E}^3}{\hat{V}^3},$$ where $\hat{T}$, $\hat{V}$ and $\hat{E}$ are the ratios of the number of triangles (\inlinegraphics{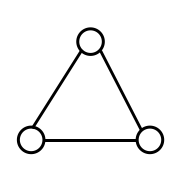}), triplets (\inlinegraphics{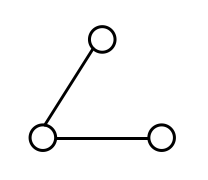}) and edges (\inlinegraphics{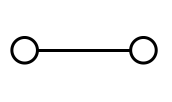}) in the network to the total number of potential connections respectively. The normalization makes the new indicator, unlike the clustering coefficient, robust to network size, network density and degree heterogeneity. Moreover, the normalized clustering coefficient reflects the "in-out-ratio" of the DCBM. In our simulations and empirical studies, the new measure out performs the clustering coefficient in networks clustering and dynamic network analysis. It is also a good criteria on measuring the quality of sub-networks in network sampling.
	
	In related work, \cite{gao2017} and \cite{jin2018} proposed similar statistics for testing the existence of communities structure based on small subgraphs. 
Partly inspired by \cite{gao2017} and \cite{jin2018}, we investigate the clustering strength of the network through three-node subgraphs, which has a quite different motivation from former works.
	
	More precisely, we make contributions as follows:
	\begin{itemize}
		\item A novel statistic that is robust to network size, network density and degree heterogeneity is proposed to measure the clustering strength of networks.
		\item The normalized clustering coefficient has nice statistical properties under different popular network generative models.
		\item Both simulations and real data analysis show that the normalized clustering coefficient can be applied for networks clustering, network sampling and dynamic network analysis.
	\end{itemize}
	
	The rest of the paper is organized as follows. We introduce the normalized clustering coefficient in detail and analyze its properties under different network generative models in Section \ref{tNCC}. In Section \ref{application}, we describe the applications of the normalized clustering coefficient in networks clustering, network sampling and dynamic networks analysis. We make conclusion in Section \ref{conclusion}. Technical proofs of the mathematical results, additional simulations and more details of the real data are given in Appendix. 
	
	\section{The normalized clustering coefficient}\label{tNCC}
	
	
	Let $G(V,E)$ denote a network with node set $V$ and edge set $E.$ Let $|V|=N.$ For any $i,j \in \{1,2,\dots,N\}$ and $i\neq j$, let $A_{i,j}=1$ if there is an edge between node $i$ and node $j;$ otherwise, $A_{i,j}=0$. Set $A_{i,i}=0$ for all $i\in \{1,2,\dots,N\}.$ The normalized clustering coefficient is defined as
	\begin{equation}\label{ncc}
	\hat{\rho}=\frac{\hat{T}\hat{E}^3}{\hat{V}^3},
	\end{equation}
	
	where 
	$\hat {E} = \binom{N}{2}^{-1} \sum_{1 \leq i \leq j \leq N}A_{i,j}$,
	$\hat{V} = \binom{N}{3}^{-1} \sum_{1\leq i\leq j\leq k \leq N}(A_{i,j}A_{i,k}+A_{j,i}A_{j,k}+A_{k,i}A_{k,j})/3$ and
	$\hat{T} = \binom{N}{3}^{-1} \sum_{1\leq i\leq j\leq k \leq N}A_{i,j}A_{j,k}A_{k,i}$.
	
	Using the adjacency matrix $A$, the normalized clustering coefficient can be rewritten as
	\begin{equation*}
	\hat{\rho} = \frac{(N-2)^2 tr(A^3)(\mathbf{1}'A\mathbf{1})^3}{N(N-1)(\mathbf{1}'A^2\mathbf{1}-tr(A^2))^3},
	\end{equation*} 
	where $tr()$ denotes the trace of the matrix and $\mathbf{1}$ is the all-one vector. This is the form we use in simulations and empirical studies.
	
	The computational complexity of the normalized clustering coefficient, which is mainly from counting triangles, is $O(Nd^2)$ (\cite{ncc_comp}), where $d$ is the maximum degree in the network. Since many networks in the real world are sparse where $d \ll N$, the complexity is modest. Furthermore, the counting of triangles and triplets can be achieved through parallel computation using the edge list, which will significantly increase the computing efficiency. 
	
    Unlike the clustering coefficient, the normalized clustering coefficient involves the number of edges and utilizes the cubic of $\hat{E}/\hat{V}$ to perform normalization. The proposed indicator has nice properties under some of the popular network generative models.

	\subsection{Under the Erd\"{o}s-R\'{e}nyi model}
Under the Erd\"{o}s-R\'{e}nyi model with link probability $p,$ the clustering coefficient
$\hat{cc}\rightarrow3p,$
and the normalized clustering coefficient
$\hat{\rho}\rightarrow1.$ The clustering coefficient was proposed to measure the degree to which nodes that share a common neighbor tend to have links. However, the Erd\"{o}s-R\'{e}nyi model assumes links appear independently, so the number of common neighbors will not change the probability that two nodes are connected. In this sense, it is more natural to conclude that the strength of this kind of connection is 1, as the normalized clustering coefficient does.
	
	\subsection{Under the DCBM}
    We first give a brief introduction of the DCBM. Then, we show that the clustering coefficient is not a proper measure while the normalized clustering coefficient has nice properties under the DCBM.
    
	In the real world, there are various sources of community structures in networks. In social networks, there are families, work and friendship circles, interest groups, universities, and companies. In protein interaction networks, proteins are often grouped according to their functions within the cells. Furthermore, nodes in real networks often show degree heterogeneity even when they are in the same community. In order to accommodate community structure as well as degree heterogeneity in networks, \cite{r6} proposed the degree-corrected stochastic block model (DCBM). The theoretical properties of DCBM have been studied for community detection, including but not limited to \cite{r4,r9, r600, r601}.
	
	The DCBM can be described as follows: given a network $G$ generated from the DCBM, each node $i \in \{1,\dots,N\}$ is assigned a class label $z_i\in \{1,\dots,K\}$ where $z_i \sim Multinomial(\cdot | \pi)$ and  $\pi=(\pi_1,...,\pi_K)$ with $\sum_{i=1}^{K}\pi_i=1$.  The adjacent matrix $A=\{A_{i,j}\}_{n \times n}$ is defined by 
	$
	A_{i,j} \sim Bernoulli(\cdot |\theta_i \theta_j B_{z_i,z_j})
	$ 
	for $i\ne j,$ and $A_{i,i}=0$ for all $i\in \{1,2,\dots,N\},$ where $\{\theta_i: i=1, \dots , N\}$ are non-negative degree-correction parameters, and $B$ is the $K\times K$ symmetric probability matrix. In general, we only require $\min\{B_{i,i}\}>\max\{B_{i,j}\},$ where $i,j\in \{1,2,\dots,K\}$ and $i\neq j.$ However, in the following discussion, we reduce the parameter domain to $B_{i,i}=p>q=B_{i,j},$ for all $i,j\in \{1,2,\dots,K\}$ and $i\neq j.$ We also set $\pi_i=\frac{1}{K}$ for all $i \in \{1,2,\dots,K\}.$ Without loss of generality, we further assume the constraint $\mathbb{E}(\theta^2) = 1$.
	
	Now, we demonstrate that the clustering coefficient is not an appropriate measure under the DCBM framework. Let $E = \mathbb{P}(A_{i,j}=1)$,
	$V = \mathbb{P}(A_{i,j}A_{i,k}=1)$ and 
	$T = \mathbb{P}(A_{i,j}A_{j,k}A_{k,i}=1)$.
	Then, under the DCBM, we have 
	\begin{eqnarray}
	E &=& (\mathbb{E}(\theta))^2\Big(\frac{p}{K}+\frac{(K-1)q}{K}\Big),  \label{EE} \\
	V &=& (\mathbb{E}(\theta))^2\Big(\frac{p}{K}+\frac{(K-1)q}{K}\Big)^2,  \label{EV} \\
	T &=& \frac{p^3}{K^2}+\frac{3(K-1)pq^2}{K^2}+\frac{(K-1)(K-2)q^3}{K^2}. \label{ET}
	\end{eqnarray}
	Hence the expectation of the clustering coefficient given in (\ref{cc}) can be approximated as:
	\begin{equation}
	\mathbb{E}(\hat{cc})\approx cc \equiv \frac{3T}{V}=\frac{ 3(p^3+3(K-1)pq^2+(K-1)(K-2)q^3) }{(\mathbb{E}(\theta))^2\big(p+(K-1)q\big)^2}.
	\end{equation}
	
	The above equation demonstrates that it is difficult to draw useful information from the clustering coefficient. First, the clustering coefficient involves $p$, $q$ and $\mathbb{E}(\theta)$, which makes it difficult to get straight forward intuition about what is the clustering coefficient measuring. Second, it will tend to $0$ as $N$ goes to infinity in sparse ($p=O(1/N)$) or semi-sparse ($p=O(\log(N)/N)$) networks. Therefore the clustering coefficient is not an appropriate measure at least under the DCBM.
	
	
	In order to modify the clustering coefficient, we aim to remove the effects of $\theta$ and network density. From (\ref{EE}) and (\ref{EV}), $\mathbb{E}(\theta)$ could be eliminated by taking the ratio $V/E.$ We divide $T$ by $(V/E)^3$ to remove link probability $p$ and obtain the normalized clustering coefficient. The expectation of the normalized clustering coefficient can be approximated as 
	\begin{equation}\label{rho}
	\mathbb{E}(\hat{\rho}) \approx \rho \equiv \frac{TE^3}{V^3}=\frac{Kr^3+3K(K-1)r+K(K-1)(K-2)}{(r+(K-1))^3},
	\end{equation}
	where $r=p/q > 1.$ Moreover, since $\frac{ d }{d r} \mathbb{E}(\rho) = \frac{3K(K-1)(r-1)^2}{(r+K-1)^4} > 0$, $\rho$ is an increasing function of $r.$ 
	Hence, the normalized clustering coefficient can be used to infer the "in-out-ratio" $r$ under the DCBM framework. The larger the value of the normalized clustering coefficient, the stronger the community effect.
	
	The next theorem shows that $\hat{\rho}$ is asymptotically normal with a high convergence rate $d^{-3/2},$ where $d=Np.$	
	\begin{theorem}\label{central}
		Under the DCBM with $K\geq 2$ as a constant, assume $\mathbb{E}(\Theta^4)=O(1)$, $N^{-1} \ll p \asymp q \ll N^{-\frac{2}{3}}$ and $\hat{V}>0.$ Let $\rho=\frac{TE^3}{V^3}.$ Then
		\begin{eqnarray}
		\frac{\sqrt{\binom{N}{3}T}(\hat{\rho}-\rho)}{\rho}\rightsquigarrow N(0,1).
		\end{eqnarray}
	\end{theorem}	
    Since $\rho \in (1, K)$ and $K$ is a given constant,  the convergence rate is $O(\sqrt{N^3T}).$ The technical proof of Theorem \ref{central} is given in Appendix A. 
	
	Since $\rho$ is a monotone function of $r$, combined with the result of Theorem \ref{central}, the value of the "in-out-ratio" of the DCBM can be inferred from the value of the normalized clustering coefficient instead of applying a community detection algorithm first and then calculating the "in-out-ratio" with the community labels. To our best knowledge, this is the first work to infer the "in-out-ratio" under the DCBM without applying community detection algorithms.
	
	\subsection{Under the LCD model}
	In this subsection we explore the normalized clustering coefficient under the LCD model, which was first introduced in \cite{rLCD}. Let $d_G(v)$ denote the degree of the vertex $v$ in the graph $G.$ Define a random graph process $(G_m^{(t)})_{t\geq 0},$ where $m$ is an integer, so that $G_m^{(t)}$ is a graph on $\{v_i : 1\leq i\leq t\}$ generated as follows: start with the graph with one vertex and $m$ self-loops $(G_m^{(1)}).$ Form $G_m^{(t)}$ based on $G_m^{(t-1)}$ by adding a new vertex $v_t$ together with $m$ edges between $v_t$ and $\{v_{i_1},v_{i_2},\dots,v_{i_m}\}$, where $i_j$ is randomly chosen with probability
	\[\mathbb{P}(i_j=k)=
	\begin{cases}
	d_{G_m^{(t-1)}}(v_k)/(2mt-1) & \quad 1\leq k \leq t-1,\\
	1/(2mt-1) & \quad k=t,
	\end{cases}
	\]
	$j \in \{1,2,\dots,m\}.$ Send $m$ edges from $v_t$ to $m$ different vertexes and the probability that a vertex is chosen is proportional to its degree. The LCD model satisfies the vague description of the generation rule from the Barab\'{a}si-Albert model (BA model), which was introduced by \cite{barabasi1999}.
	
	The normalized clustering coefficient has the following property under the LCD model:
	\begin{theorem}\label{tLCD_ncc}
		Let $m \geq 1$ be fixed. Then as $N\rightarrow \infty,$ we have
		$$\mathbb{E}(\hat{\rho})\sim \frac{3m(m-1)}{4(m+1)^2}.$$
	\end{theorem}
	From Theorem \ref{tLCD_ncc}, the normalized clustering coefficient only depends on the parameter $m.$ Furthermore, since $\frac{3m(m-1)}{4(m+1)^2}$ is a monotone function of $m,$ we can infer $m$ from the value of the normalized clustering coefficient. The proof of Theorem \ref{tLCD_ncc} is given in Appendix A.
	
\begin{figure}[h!]
		\centering
		\includegraphics[scale=0.3]{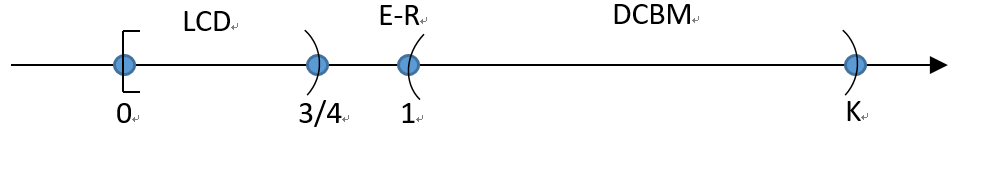}\caption{}\label{diff}
	\end{figure}

	Investigating the expectation of the normalized clustering coefficient, it is interesting to find that its range is $[0,\frac{3}{4})$ under the LCD model, 1 under the Erd\"{o}s-R\'{e}nyi model and $(1,K)$ under the DCBM as shown in Figure \ref{diff}. Therefore, we can roughly identify the generative model of the observed network by simply counting the normalized clustering coefficient.
		
	\section{Applications}\label{application}
 In this section, we describe several possible applications of the normalized clustering coefficient in network data analysis.	
	
	\subsection{Networks clustering}\label{Classification}
Given two networks $G_1$ and $G_2$ with adjacency matrices $A_1$ and $A_2,$ a natural question is: are they generated from the same probabilistic generative model? This is referred to as networks clustering, which attempts to group similar networks together. In the computer science literature (\cite{rnet_kern_she}, \cite{rnet_kern_wolf}), kernel based similarity measures between networks are proposed and computed efficiently. Recently, two algorithms called Networks Clustering based on Graphon Estimates (NCGE) and Networks Clustering based on Log Moments (NCLM) are proposed by \cite{rnet_nips} to deal with the networks clustering task with or without node correspondence.

In this subsection, we demonstrate that the normalized clustering coefficient could provide important information about the structure of networks, and thus is potentially useful for networks clustering. Intuitively, we can compare the normalized clustering coefficients, $\hat{\rho_1}$ and $\hat{\rho_2},$ of $G_1$ and $G_2$ respectively. If $|\hat{\rho_1}-\hat{\rho_2}|$ is larger than a certain threshold $c,$ we will reject the claim $H_0$: $G_1$ and $G_2$ are generated from the same generative model.

\subsubsection{A real example: Twitter network}
\begin{figure}[h!]
	\centering
	\begin{subfigure}[b]{0.4\linewidth}
		\includegraphics[width=\linewidth]{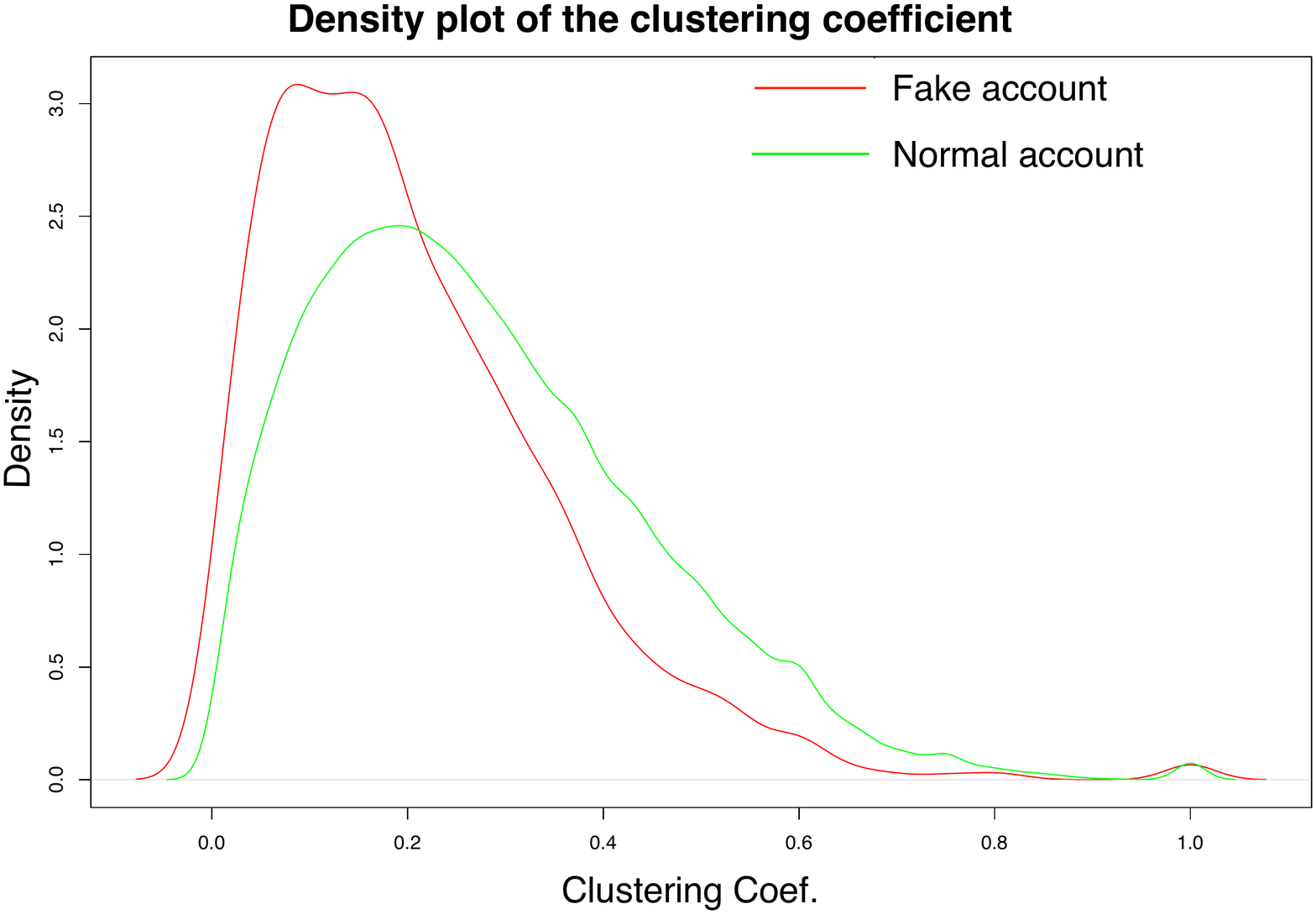}
		\caption{ }
	\end{subfigure}
	\begin{subfigure}[b]{0.4\linewidth}
		\includegraphics[width=\linewidth]{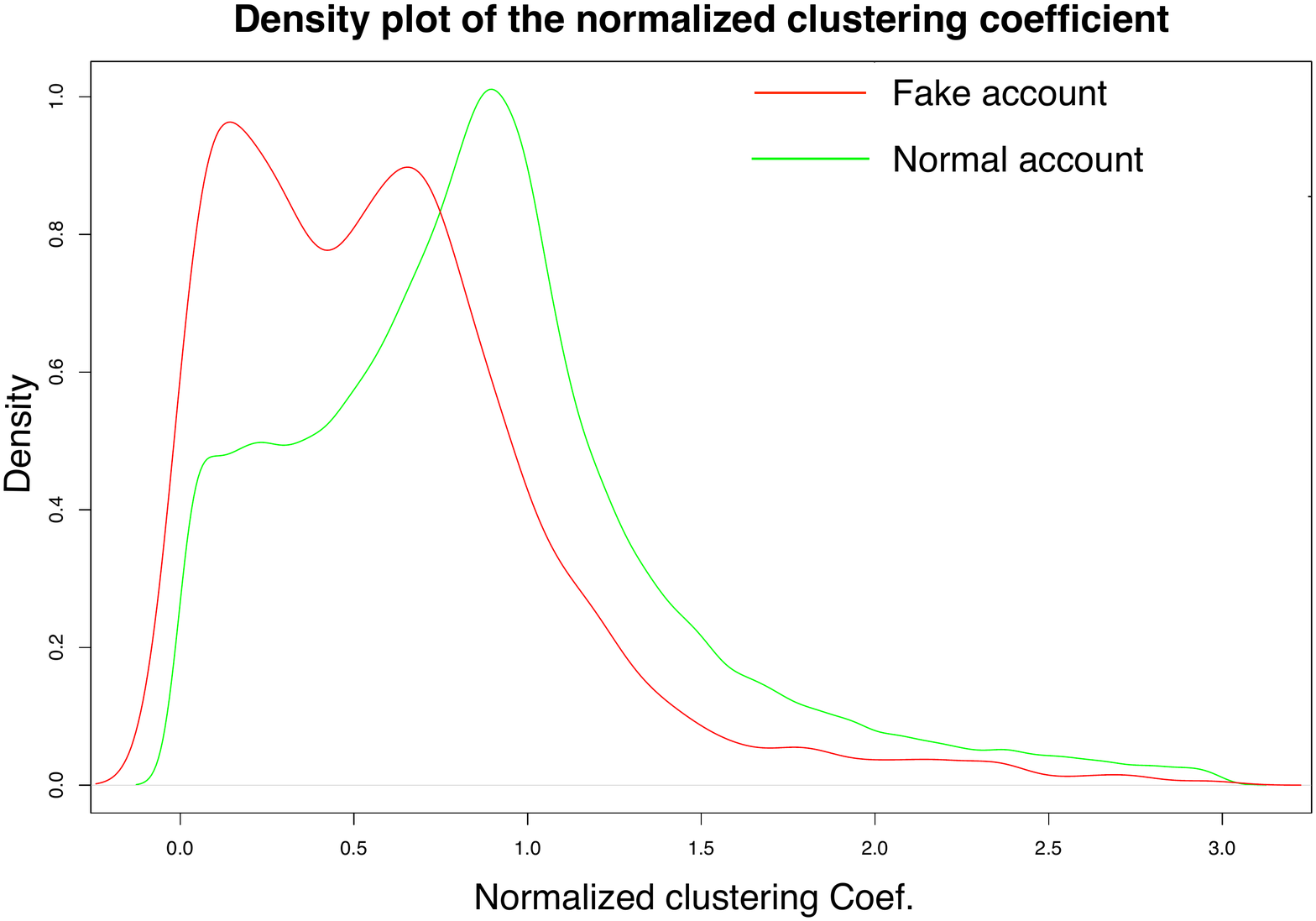}
		\caption{}
	\end{subfigure}		\begin{subfigure}[b]{0.4\linewidth}
		\includegraphics[width=\linewidth]{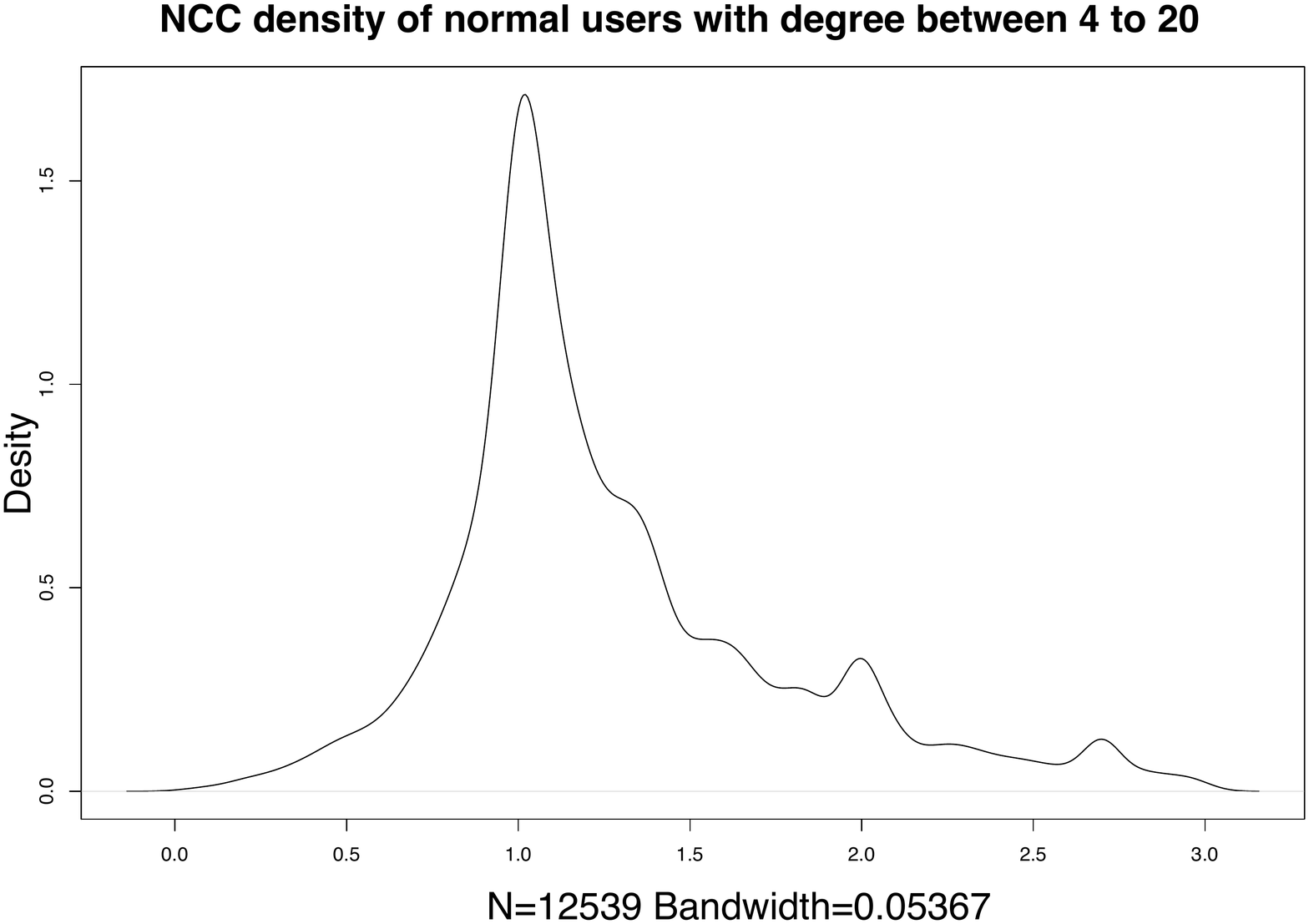}
		\caption{}
	\end{subfigure}
	\begin{subfigure}[b]{0.4\linewidth}
		\includegraphics[width=\linewidth]{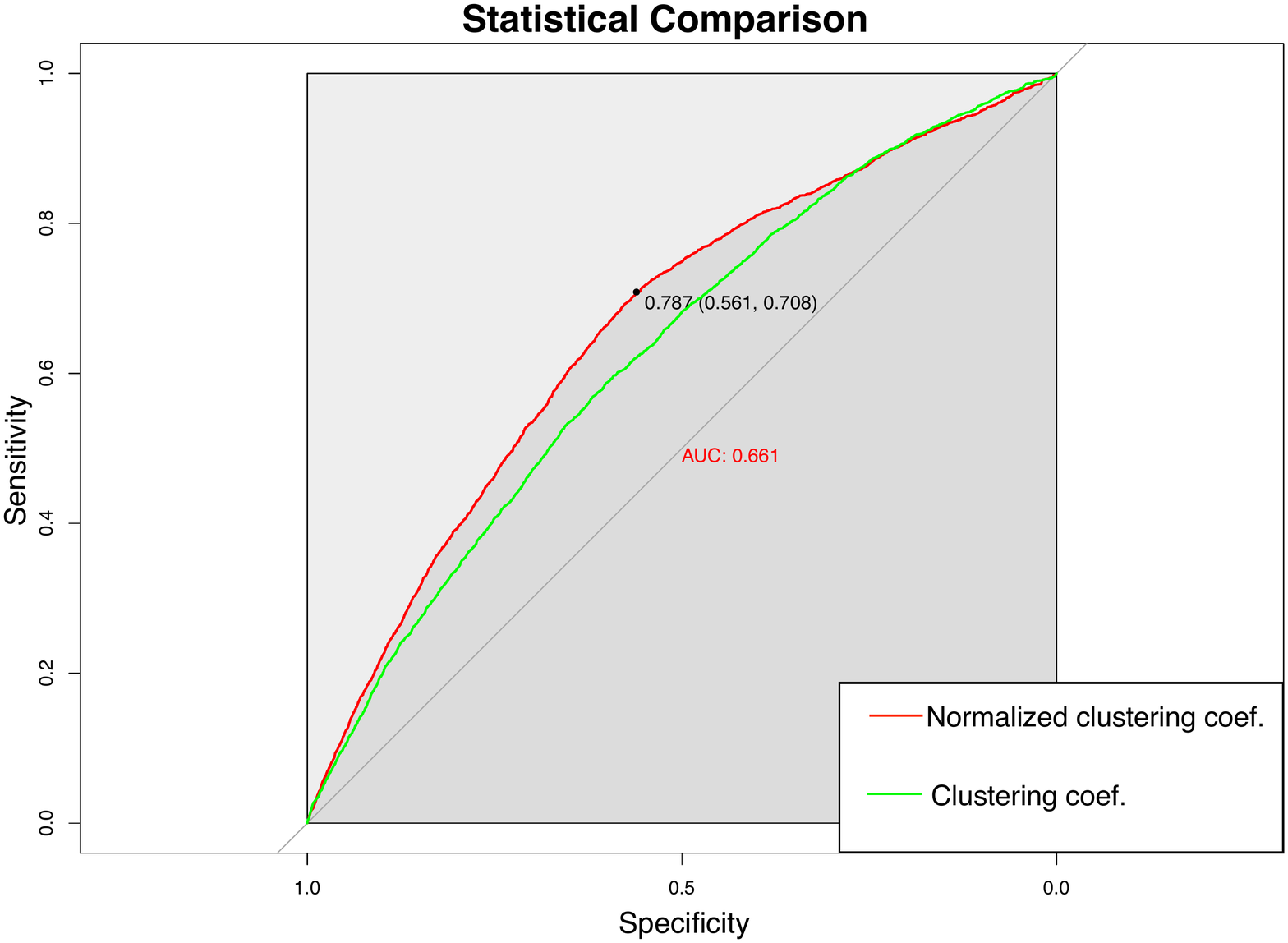}
		\caption{}
	\end{subfigure}
	\caption{(a)Density plot of the clustering coefficient. (b)Density plot of the normalized clustering coefficient. (c)Density plot of the normalized clustering coefficient of low-degree users. (d)The ROC curves.}\label{twitter_density}
\end{figure}

 We calculate the normalized clustering coefficient of Twitter ego networks to detect abnormal accounts. More details about the dataset and data preprocessing can be found in \cite{rnet_real} and Appendix B. The density plot of the normalized clustering coefficient and the clustering coefficient are shown in Figure \ref{twitter_density}.

From Figure \ref{twitter_density}(a), the distribution of the clustering coefficient of fake accounts (red line) is very similar to that of normal accounts (green line). However, the distributions of the normalized clustering coefficient of the two types of accounts (Figure \ref{twitter_density} (b)) are quite different. The normal accounts have larger normalized clustering coefficients than the fake accounts, which is consistent with the reality that there are community structures in real social networks while the fake users create links randomly. 

It is also interesting to see that the density curve of the normal accounts' normalized clustering coefficient has a small peak between 0 and 0.5. This might be because the data are not completely labeled, which means that many fake accounts are incorrectly treated as normal ones, and most of them have small normalized clustering coefficients. This represents evidence of the claim in \cite{rnet_real} that Twitter operators have not discovered a substantial number of malicious accounts. In order to test our assumption, we randomly sampled 50,000 normal users whose degree is between 4 and 20, and we believe few of them are unlabeled fake users since fake users tend to connect to many users. We plot the density of their normalized clustering coefficient which is larger than 0 in Figure \ref{twitter_density}(c). The plot confirms that without unlabeled fake users, the distribution of the normalized clustering coefficient is normal. 

Even with labeling issues, the normalized clustering coefficient still out performs the clustering coefficient. We calculate the ROC curves of these two statistics for comparison in \ref{twitter_density}(d). The AUC of the normalized clustering coefficient is 0.661 (red) and that of the clustering coefficient is 0.627 (green). The optimal threshold of the normalized clustering coefficient is 0.787 with specificity 0.561 and sensitivity 0.708.

It would be more flexible if we treat the normalized clustering coefficient as a new feature of networks and add it into other networks clustering algorithms. More simulation results of networks clustering can be found in Appendix B.

\subsubsection{Theoretical result for the DCBM}

To illustrate the above idea, we look at a special case of the DCBM with the same number of groups and balanced group size. Then the networks clustering problem is reduced to testing whether two networks have the same "in-out-ratio" for their generative models. 
	
\begin{Proposition}\label{test_dcbm}
Let $G_{1}$ be a network generated from the DCBM with parameters $\{N_{1}, B, K, \pi, p_{1}, q_{1}, \Theta_1\}$, and $G_{2}$ be a network generated from the DCBM with parameters $\{N_{2}, B, K, \pi, p_{2}, q_{2},\Theta_2\}$, where $\pi_{i}=1/K, i\in \{1,\dots, K\}.$
Let $r_{1}=p_{1}/q_{1}$ and $r_{2}=p_{2}/q_{2}.$ Given adjacency matrices $A_1$ and $A_2$ of $G_1$ and $G_2$ respectively, we want to test
$$H_{0}: r_{1} = r_{2},$$
$$H_{1}: r_{1} \neq r_{2}.$$	
\end{Proposition}

	Here, we allow the size of networks, the average degree of networks as well as the degree distribution of nodes to be different, which reflects the situation in the real world.
	
	 From Theorem \ref{central}, we have $(\hat{\rho}_1-\hat{\rho}_2)\sim N(\rho_1-\rho_2, \frac{\rho_1^2}{\binom{N_1}{3}T_1}+\frac{\rho_2^2}{\binom{N_2}{3}T_2}).$ Under null hypothesis, we obtain $(\hat{\rho}_1-\hat{\rho}_2)\sim N(0, \frac{\rho^2}{\binom{N_1}{3}T_1}+\frac{\rho^2}{\binom{N_2}{3}T_2}).$ Notice that in practice, the value of $\rho$ is unknown. However, from equation (\ref{rho}), we have $\rho \in (1,K).$ Therefore, as long as the average degree is $\omega(1)$, we can construct the following powerful test based on the normalized clustering coefficient:
	\begin{Proposition}\label{test_ncc}
		To test Proposition \ref{test_dcbm}, we compute the normalized clustering coefficient of each network to obtain $\hat{\rho}_1$ and $\hat{\rho}_2$ first. We reject the null hypothesis when $|\hat{\rho}_1-\hat{\rho}_2|>\phi^{-1}(1-\frac{\alpha}{2}){K\sqrt{6}}{\sqrt{\frac{1}{d_1^3}+\frac{1}{{d_2^3}}}}$ with significant level $\alpha,$ where $d_1$ and $d_2$ are the average degrees of two networks.
	\end{Proposition}
	Now, let us consider the power of the test in Proposition \ref{test_ncc}.
	
	\begin{theorem}\label{test_thm}
		Assume $\mathbb{E}(\Theta_1^4)=\mathbb{E}(\Theta_2^4)=O(1)$, $N_1=O(N)$, $N_2=O(N)$ and $N^{-1}_i \ll p_i \asymp q_i \ll N^{-\frac{2}{3}}_i,$ where $i \in \{1,2\}$ and $\Delta \rho= |\rho_1-\rho_2|=\omega(d^{-\frac{3}{2}}),$ where $d=\min\{d_1, d_2\}.$ Then,
		$$\mathbb{P}\left(Reject \  H_0|H_1\right)\rightarrow 1.$$
	\end{theorem}
	Although we have assumed the number of groups to be constant, it can be generalized to $K=o(d^\frac{3}{2}),$ when $\Delta \rho = O(1).$ The proof of Theorem \ref{test_thm} is given in Appendix A.
	
	\subsection{Network sampling}\label{Subsampling}
	In the real world, the enormous number of network nodes makes it infeasible to study the entire network computationally or visually. 
Therefore, network sampling is applied to understand the network structure and visualize the network.
	
	Many sampling schemes have been proposed in the literature. The most intuitive one must be random node sampling, which chooses nodes independently and uniformly at random from the original network. However, the scheme is shown in \cite{lee2005} that due to its inclusion of all the edges for a sampled node set only, it is less likely to preserve the original level of connectivity. Several variations of node sampling have been proposed in the literature (for instance, \cite{leskovec2006}; \cite{ahmed2014}). Other sampling methods such as random walk based sampling and its derivatives and snow ball sampling are also widely used in practice. We recommend \cite{rezvanian2019} for an excellent review of network sampling.
	
	With so many sampling schemes to choose from, it is crucial to ask which one to pick, and more importantly, how to evaluate the sample quality. In \cite{leskovec2006}, novel methods were proposed to check the goodness of sampling. In this section, we will show that the normalized clustering coefficient is a good criterion to evaluate the goodness of sampling.
	
	First, let us state the network sampling in mathematical form. Given an input graph $G(V,E)$ and a sampling fraction $f$, a sampling algorithm samples a representative sub-graph $G_s(V_s,E_s)$ with a subset of the nodes $V_s \subset V$ and a subset of the edges $E_s \subset \{e_{ij}|v_i\in V_s,v_j \in V_s\}$, such that $|V_s|=N_s=fN$. The goal is to ensure that the sampled sub-graph $G_s$ preserves the properties of the original graph $G$. Let the original network generated from the DCBM have parameters $\{N, K, \pi, p, q, \Theta\}.$ We want $G_s$ to have the same properties as a network generated from the DCBM with parameters $\{N_s, K_s, \pi_s, p_s, q_s, \Theta_s\}.$ It is hard to keep $p_s=p$ and $q_s=q$ because the average degree will probably change after sampling. However, it is reasonable to require $r_s=p_s/q_s=p/q=r.$ Recall equation \ref{rho}, it is a necessary condition that $\rho_s=\rho,$ which implies $\hat{\rho}_s$ should be close to $\hat{\rho}.$ Since we cannot make sure $K_s=K$ and $\pi_s=\pi,$ the closeness of the normalized clustering coefficient between the sub-network and the original network is not a sufficient condition to ensure the sub-network preserves the structure of the original one. However, it can help to screen out the bad sampling results and pick out the best ones.
	
	In order to evaluate popular network sampling methods, we use three huge networks from different domains, the Arxiv COND-MAT (condense matter physics) collaboration network, the Enron email network and a human protein-protein interaction network (all are downloaded from \url{http://snap.stanford.edu/}). We apply random node sampling (NS), random edge sampling (ES), random walk sampling (RWS), random walk with flying back sampling (RWFS), random walk with jumping out sampling (RWJS), forest fire sampling (FF) and snowball sampling (SS) to draw 10 sub-networks each with sampling fraction $f=10\%$. More details of the data and sampling methods are given in Appendix C. We calculate the normalized clustering coefficient of each sub-network and obtain the mean and standard deviation for each sampling method to compare them with the original normalized clustering coefficient. The results are shown as follows:
		\begin{center}
		\scalebox{0.85}{\begin{tabular}{ |c c c c c c c c c| } 
			\hline
			& Original &NS &ES &RWS &RWJS &RWFS &FF &SS   \\ 
			\hline 
			ConMatter &103.68 &123.52\tiny{(43.02)} &17.83\tiny{(1.46)} &24.91\tiny{(2.99)} &37.50\tiny{(5.56)} &20.81\tiny{(4.90)} &11.58\tiny{(2.19)} &12.35\tiny{(3.85)} \\
			\hline 
			Email &1.54 &1.80\tiny{(1.43)} &1.91\tiny{(0.02)} &1.99\tiny{(0.06)} &1.93\tiny{(0.04)} &1.92\tiny{(0.10)} &1.11\tiny{(0.18)} &0.92\tiny{(0.43)}\\
			\hline 
			PPI & 2.53 &2.89\tiny{(1.43)} &2.01\tiny{(0.06)} &2.08\tiny{(0.09)} &1.88\tiny{(0.07)} &1.98\tiny{(0.08)} &1.35\tiny{(0.49)} &1.05\tiny{(0.51)}\\
			\hline 
      
		\end{tabular}}
	\end{center}
	
	\begin{figure}[h]
		\centering
		\begin{subfigure}[b]{0.3\linewidth}
			\includegraphics[width=\linewidth]{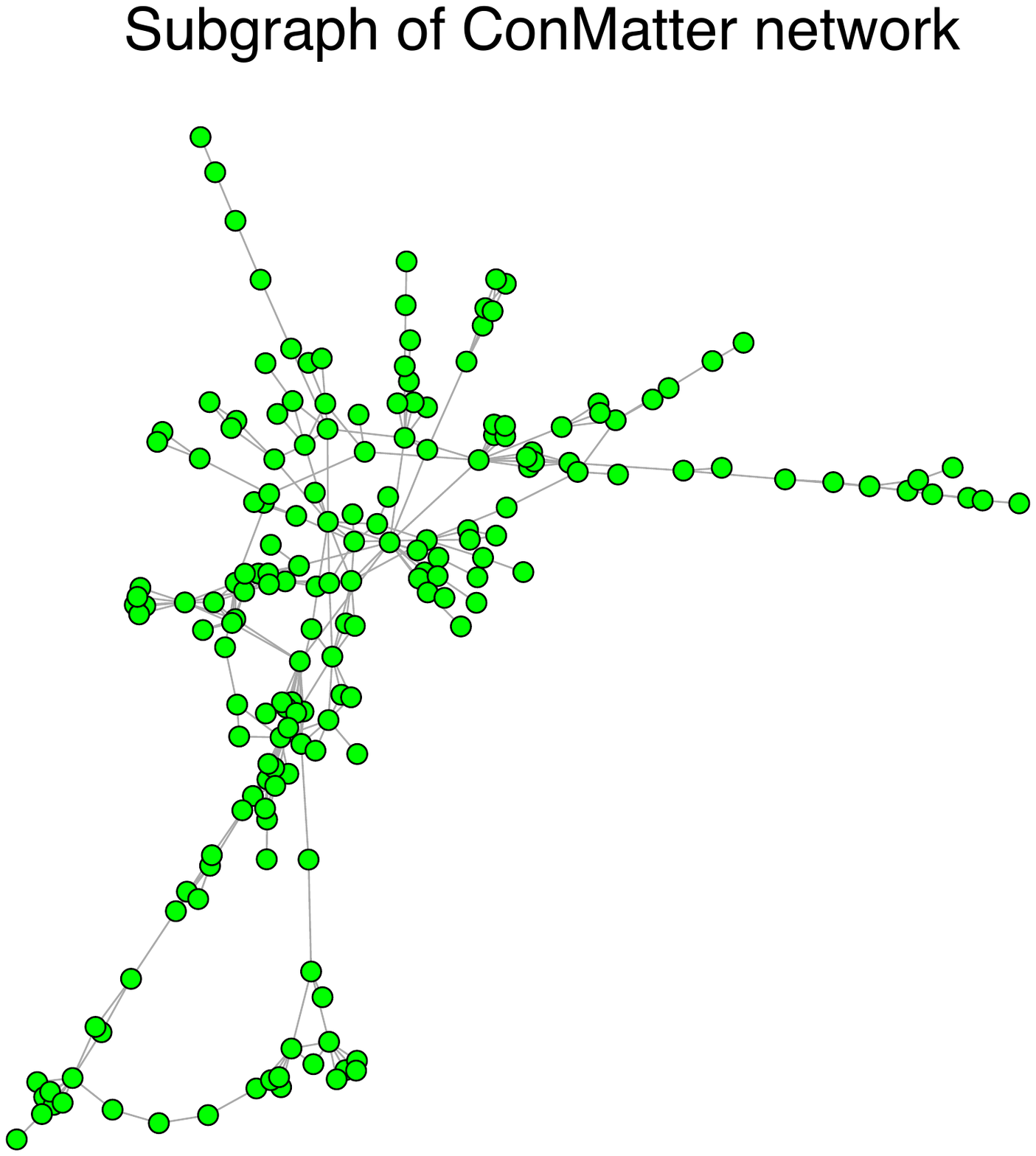}
			\caption{Sub-graph of ConMatter }
		\end{subfigure}
		\begin{subfigure}[b]{0.3\linewidth}
			\includegraphics[width=\linewidth]{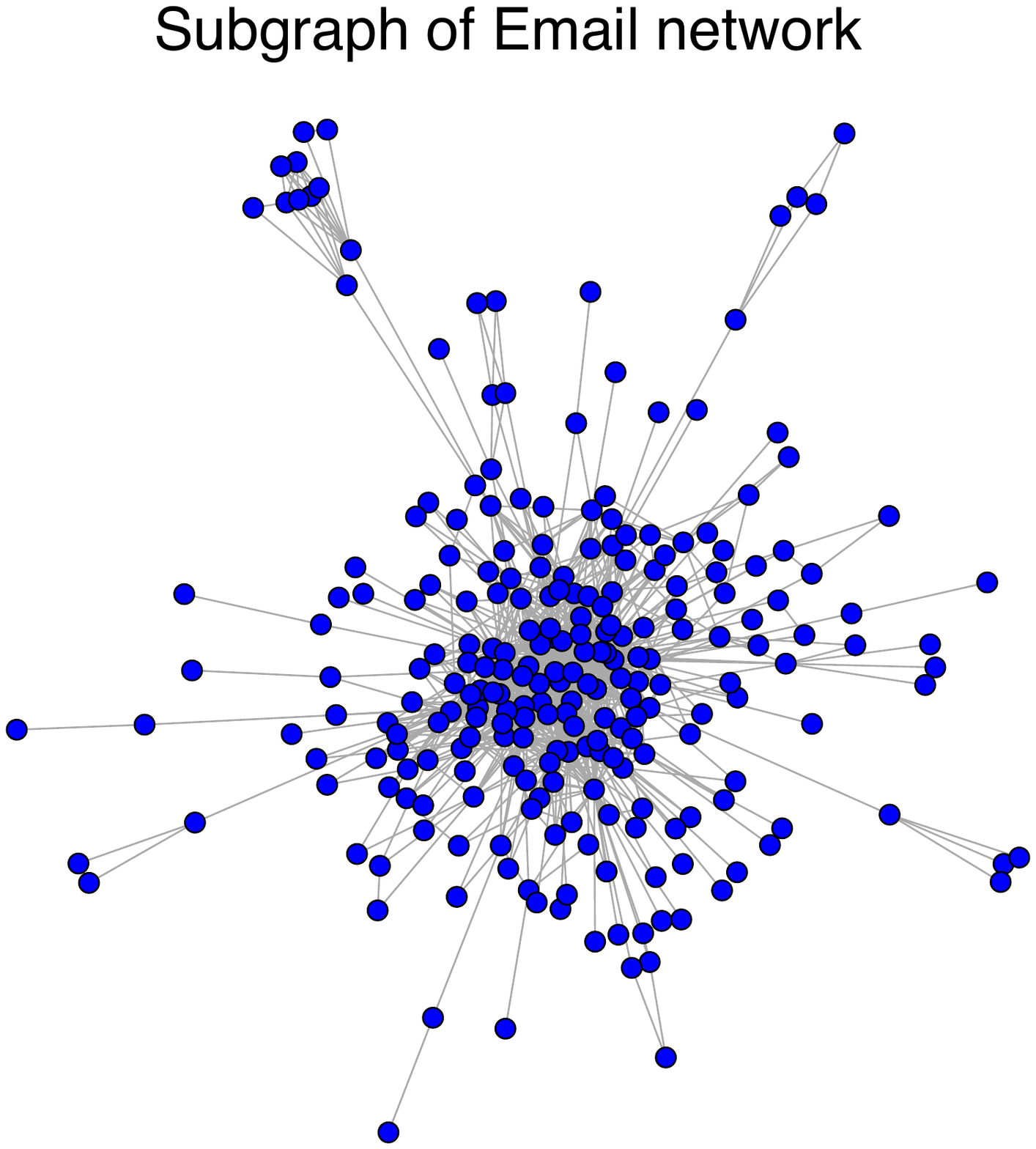}
			\caption{Sub-graph of Enron network }
		\end{subfigure}
		\begin{subfigure}[b]{0.3\linewidth}
			\includegraphics[width=\linewidth]{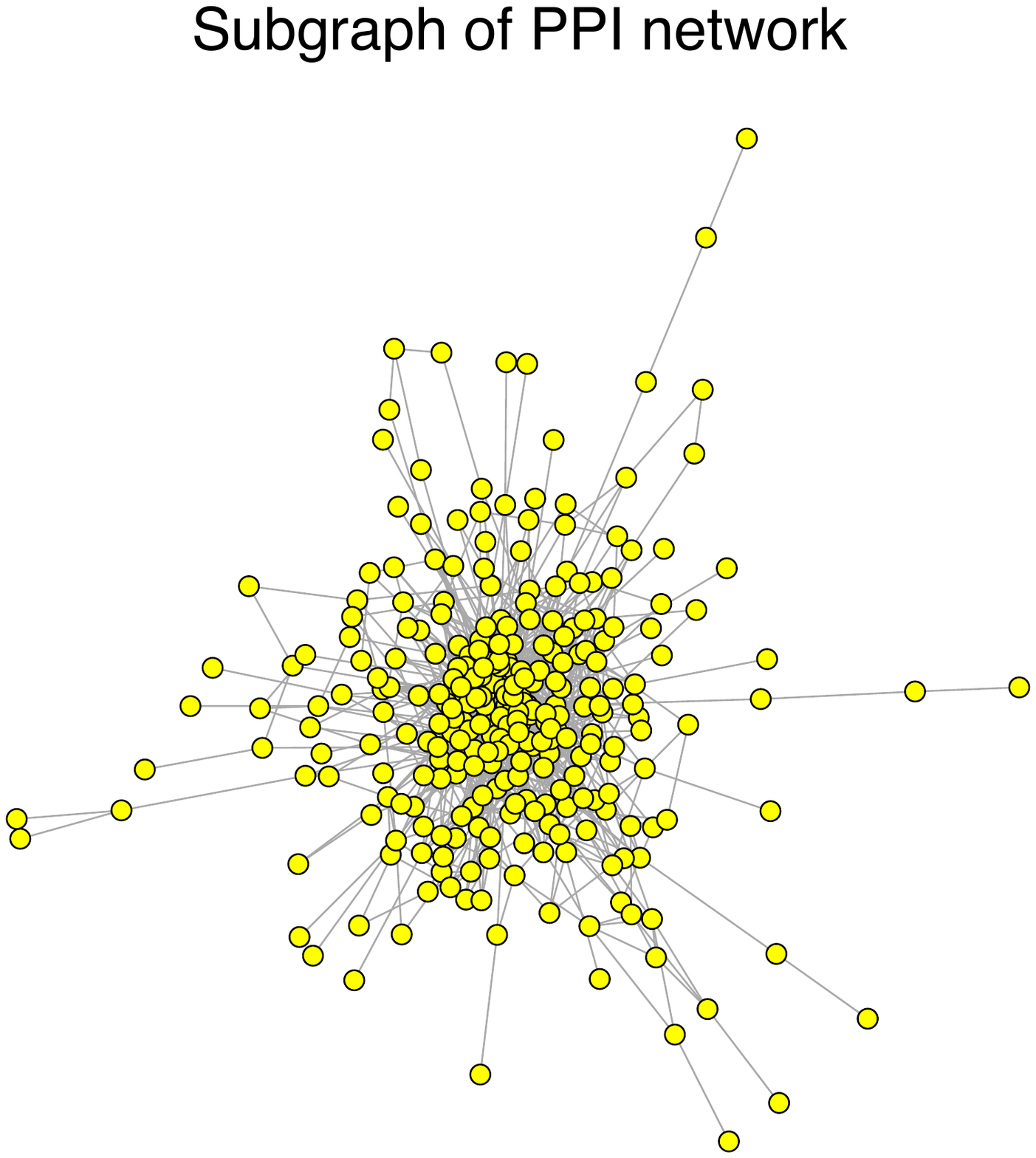}
			\caption{Sub-graph of PPI network}
		\end{subfigure}
		\caption{ }\label{subnetworks}
	\end{figure}		
%
%
%
%
%
%
	From the above table, we find that none of the sampling methods is always the best for every dataset. NS has quite large standard deviation because it often picks sub-networks with many isolated nodes. 
The disconnected issue is more serious with NS when the original network is sparse. All the methods fail on the co-author network. This is mainly because there are too many communities in the network (the original normalized clustering coefficient is 103.68, which indicates the number of communities should be larger than 104), and it is difficult to cover all the communities in the sub-network with a limited size. In other words, it is hard to achieve $K_s=K$, so $\hat{\rho}_s$ will surely be far from $\hat{\rho}.$ However, this reflects the informativeness of the normalized clustering coefficient for network sampling from another point of view. 

All the random walk based methods (RWS, RWJS and RWFS) have relatively good and stable performance. FF and SS are not so good mainly because of the bias they introduced. Since both FF and SS strongly rely on the initial node, they tend to sample the nodes from the same community with the initial node,which will lead to inaccurate estimation of the community proportion ($\pi$). 

We plot the sub-network whose normalized clustering coefficient is the closest to that of the original large network in Figure \ref{subnetworks} to visualize the structure of the original huge network.

More network sampling simulations can be found in Appendix C.
	
	\subsection{Dynamic network analysis}\label{Dynemic}
	
	Networks often evolve with time. It is of great interest to capture the changing dynamics. Some research has focused on the problem of change point detection. We recommend \cite{rchangepoint} for a nice review. Detecting communities in dynamic networks is also studied both in statistics and computer science and is well reviewed in \cite{rcddn}.
	
	When performing analysis on dynamic networks, computational capability is required not only for large graphs, but for a long sequence of large graphs over time. The computational complexity of the algorithm is crucial especially for online learning over a short time. The normalized clustering coefficient, which only requires simple counting of triangles, could be updated every second, so it is helpful in monitoring the rapid revolution of dynamic networks. It is also useful for evaluating the clustering effect and is robust to changes in the number of vertexes and average degree.
	
	We test the performance of the normalized clustering coefficient on the co-sponsorship networks structured from bills sponsored in the U.S. Senate during the 93rd-114th Congresses (\cite{rsenate_data}). More details of the data are given in Appendix D. We compute the normalized clustering coefficient, the clustering coefficient and the true "in-out-ratio" for each year's network. The true "in-out-ratio" is computed as
	$\frac{\mbox{number of edges  between congress-persons in the same party}}{\mbox{number of edges between congress-persons in different parties}}.$
	
	\begin{wrapfigure}{r}{7cm}
		\centering
	    \vspace{-15pt}
		\includegraphics[width=0.7\linewidth]{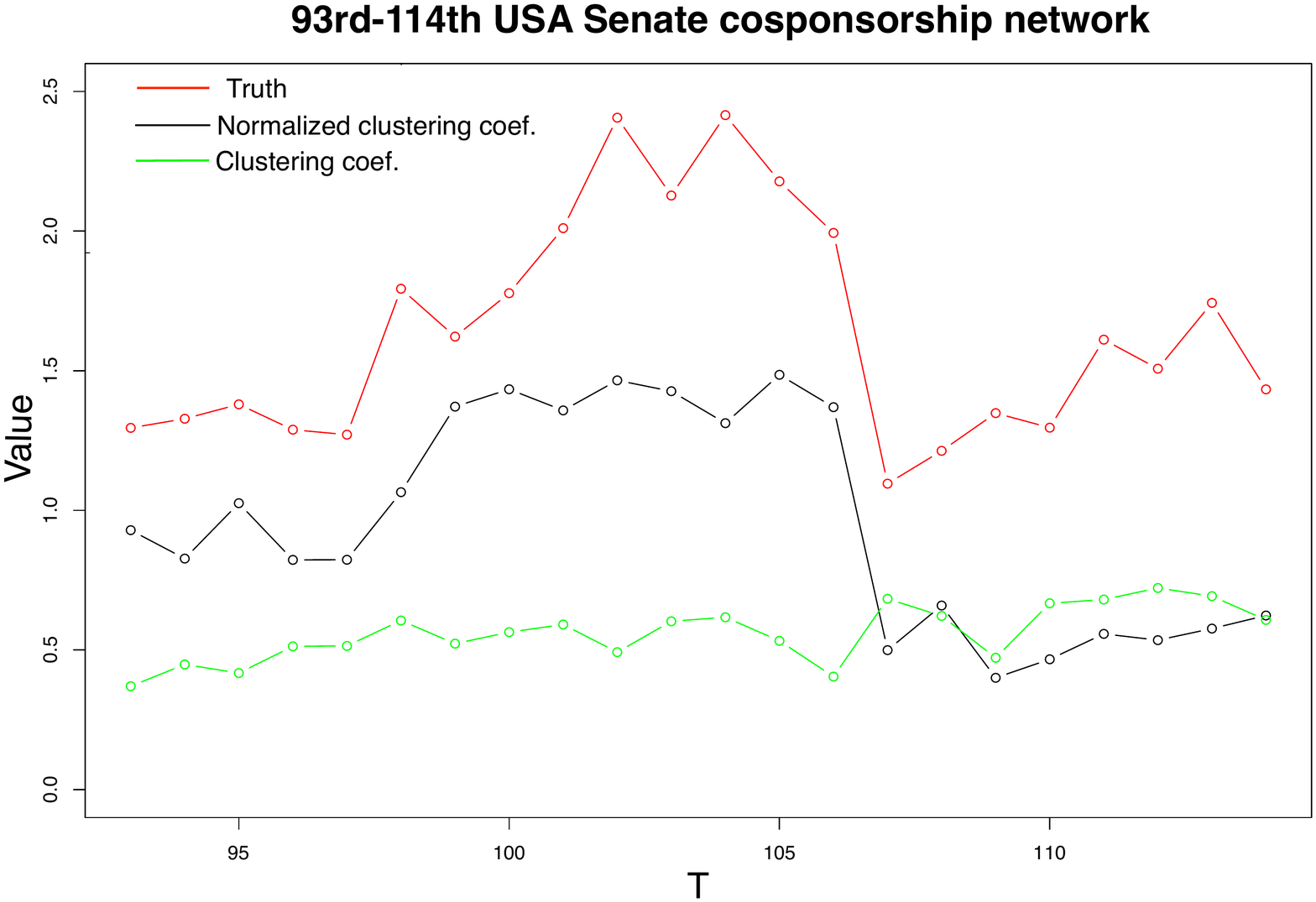}
		\caption{{\small
				Plot of the clustering strength of the co-sponsorship networks structured from bills sponsored in the U.S. Senate during 93rd-114th Congresses. The normalized clustering coefficient (black line) can reveal the truth (red line) without using the party labels, while the clustering coefficient (green line) cannot.
		}} 
		\label{senate_us}
	    \vspace{-30pt}
	\end{wrapfigure}
	
	Figure \ref{senate_us} illustrates that the normalized clustering coefficient can infer the true community strength without using the party labels, while the clustering coefficient cannot. There is an obvious change point during the 107th Senate. During that senate (January 3, 2001 to January 3, 2003), a rare even split occurred and the defection of a single senator led to three changes in majorities. The 9-11 attacks were highly disruptive and the U.S. Congress voted to allow the president to invade Iraq. The same attitudes from the two parties toward these events may explain the low community strength in the Senate.

	\section{Conclusion}\label{conclusion}
	We proposed a novel network statistic, the normalized clustering coefficient, to measure the clustering effect, and it is robust to the degree heterogeneity, network size and network density. The statistic not only has nice statistical properties, but its computational complexity is modest which makes it suitable for extremely large networks and rapidly varying dynamic networks. The statistic can also be applied for networks clustering, network sampling and dynamic network analysis.

\end{document}



\maketitle

\section*{Appendix A} 
\subsection*{Proof of Theorem 1}
The following lemmas, which will be used later, are Lemma C.1 and Lemma C.2 from \cite{gao2017}:
\begin{Lemma}\label{lemma_V}
	Assume $\mathbb{E}\Theta^4=O(1)$ and  $N^{-1}\ll p \asymp q=o(1).$ Then
	$$\mathbb{E}(\hat{E}-E)^2=O(\frac{p^2}{n}), \ \ and \ \  \mathbb{E}(\hat{V}-V)^2=O(\frac{p^4}{n}).$$
\end{Lemma}

\begin{Lemma}\label{T}
	Assume $\mathbb{E}\Theta^4=O(1)$ and  $N^{-1}\ll p \asymp q=N^{-\frac{2}{3}}.$ Then
	$$\mathbb{E}(\hat{T}-T)^2 \asymp O(\frac{p^3}{N^3}), \ \ and \ \  \frac{\sqrt{N \choose 3}(\hat{T}-T)}{\sqrt{T}}\rightsquigarrow N(0,1).$$
\end{Lemma}

\textit{Proof of Theorem 1.} 
\begin{eqnarray*}
	\frac{\hat{T}\hat{E}^3}{\hat{V}^3}
	&=&(\hat{T}-T+T)\left(\frac{\hat{E}}{\hat{V}}-\frac{E}{V}+\frac{E}{V}\right)^3\\
	&=& \frac{TE^3}{V^3}+(\hat{T}-T)\frac{E^3}{V^3}+3\hat{T}\ \left( \frac{E}{V}\right) ^2\frac{\hat{E}-E}{V}+ \\
	&&3\hat{T}\ \left( \frac{E}{V}\right) ^2\left(\frac{1}{\hat{V}}-\frac{1}{V}\right)E+3\hat{T}\ \left( \frac{E}{V}\right) ^2\left(\frac{1}{\hat{V}}-\frac{1}{V}\right)(\hat{E}-E)+\\
	&&3\hat{T}\ \left( \frac{E}{V}\right) \left(\frac{\hat{E}}{\hat{V}}-\frac{E}{V}\right)^2+3\hat{T} \left(\frac{\hat{E}}{\hat{V}}-\frac{E}{V}\right)^3.
\end{eqnarray*}
By Lemma \ref{lemma_V} and Lemma \ref{T}, $ \frac{TE^3}{V^3}+(\hat{T}-T)\frac{E^3}{V^3}$ is the dominating term. Hence
$$\hat{\rho} \rightsquigarrow N(\rho, \frac{\rho^2}{{N \choose 3}T}),$$
which is the result.

\subsection*{Proof of Theorem 2}
The following two lemmas, which are Theorem 14 and Theorem 16 from \cite{rLCD_result}, describe the number of triangles and triplets in $G_m^{(t)}.$
\begin{Lemma}\label{LCD_T}
	Let $m\geq 1$ be fixed. The expected value of the number of triangles in $G_m^{(t)}$ is given by 
	$$(1+o(1))\frac{m(m-1)(m+1)}{48}(\log N)^3$$
	as $N\rightarrow \infty.$
\end{Lemma} 

\begin{Lemma}\label{LCD_V}
	Let $m\geq 1$ and $\epsilon > 0$ be fixed. Then  
	$$(1-\epsilon)\frac{m(m+1)}{2}N\log N\leq {N \choose 3}V\leq (1+\epsilon)\frac{m(m+1)}{2}N\log N$$
	holds with a high probability as $N\rightarrow \infty.$
\end{Lemma} 

For a graph $G_m^{(t)}$ generated from the LDC model, the number of edges is fixed at $mN.$ Combining this fact with Lemma \ref{LCD_T} and Lemma \ref{LCD_V}, we can easily reach the result in Theorem 2.

\subsection*{Proof of Theorem 3}
Similar to the proof of Theorem 1, 
$\hat{\rho}_1-\hat{\rho}_2$ is dominated by 
$$\rho_1-\rho_2+\frac{\rho_1}{{N_1 \choose 3}T_1}\frac{\sqrt{N_1 \choose 3}(\hat{T}_1-T)}{\sqrt{T_1}}-\frac{\rho_2}{{N_2 \choose 3}T_2}\frac{\sqrt{N_2 \choose 3}(\hat{T}_2-T)}{\sqrt{T_2}}.$$ 
From Lemma \ref{T} we have 
$$\hat{\rho}_1-\hat{\rho}_2 \rightsquigarrow N(\rho_1-\rho_2,\frac{\rho_1^2}{{N_1 \choose 3}T_1}+\frac{\rho_2^2}{{N_2 \choose 3}T_2}).$$
Without loss of generality, let $r_1 > r_2.$ Then $\rho_1 > \rho_2.$ 
Therefore, from the assumption, we have
\begin{eqnarray*}
\mathbb{P}\left(Reject \ H_0 |H_1\right)
&\geq&
\mathbb{P}\left(d^{\frac{3}{2}}|\hat{\rho}_1-\hat{\rho}_2|\geq c |H_1\right) \ \left(here \ c=\phi^{-1}(1-\frac{\alpha}{2}){K\sqrt{6}}\right)\\ 
&\geq& 1-\mathbb{P}\left(d^{\frac{3}{2}}|\hat{\rho}_1-\hat{\rho}_2|\leq c |H_1\right)\\
&\geq & 1-\mathbb{P}\left(d^{\frac{3}{2}}(\hat{\rho}_1-\hat{\rho}_2)<c |H_1\right)\\
&\geq& 1-\mathbb{P}\left(d^{\frac{3}{2}}(\Delta \rho-\hat{\rho}_1+\hat{\rho}_2)\geq (d^{\frac{3}{2}}\Delta \rho - c) |H_1\right)\\
&\geq &
1-\frac{1}{1+\frac{(\Delta \rho-cd^{-\frac{3}{2}})^2}{\frac{\rho_1^2}{{N_1 \choose 3}T_1}+\frac{\rho_2^2}{{N_2 \choose 3}T_2}}} \ (by \ Cantelli \ inequality)\\
&\geq &
1-\frac{1}{1+\frac{\Delta \rho^2d^3}{\rho_1}}\\
& \geq &
1-\frac{1}{1+\frac{\Delta \rho^2d^3}{K}}\\
& \geq &
1-\frac{1}{1+\omega(\frac{1}{K})}\\
&\rightarrow & 1,
\end{eqnarray*}
which is the result.

\section*{Appendix B} 
\subsection*{Simulations on networks clustering}
Simulation studies are conducted to compare the performance of the clustering coefficient and the normalized clustering coefficient in networks clustering. 

\begin{figure}[H]
	\centering
	\begin{subfigure}[b]{0.45\linewidth}
		\includegraphics[width=\linewidth]{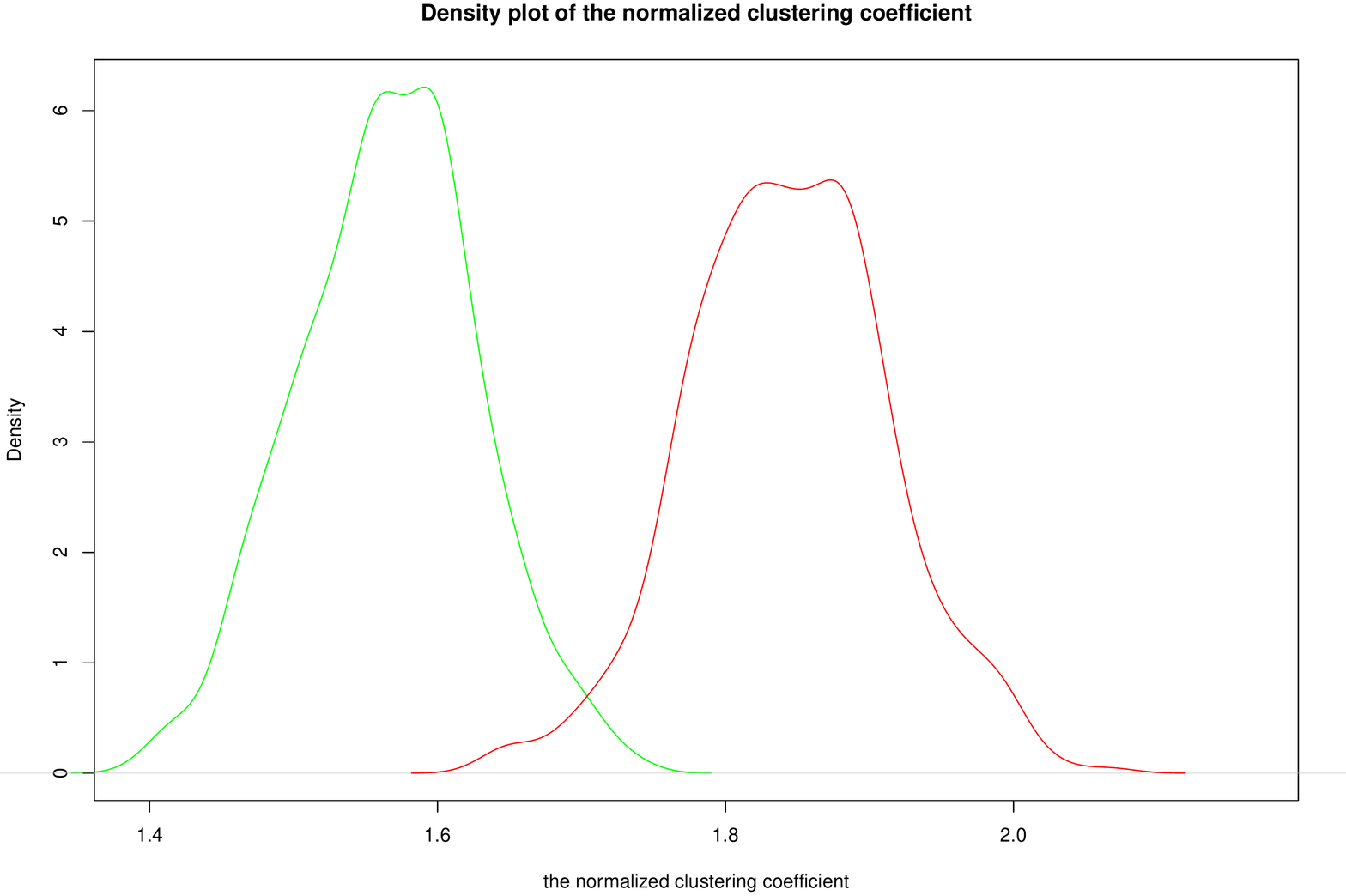}
		\caption{Density plot of the normalized clustering coefficient in Simulation 1.}
	\end{subfigure}
	\begin{subfigure}[b]{0.45\linewidth}
		\includegraphics[width=\linewidth]{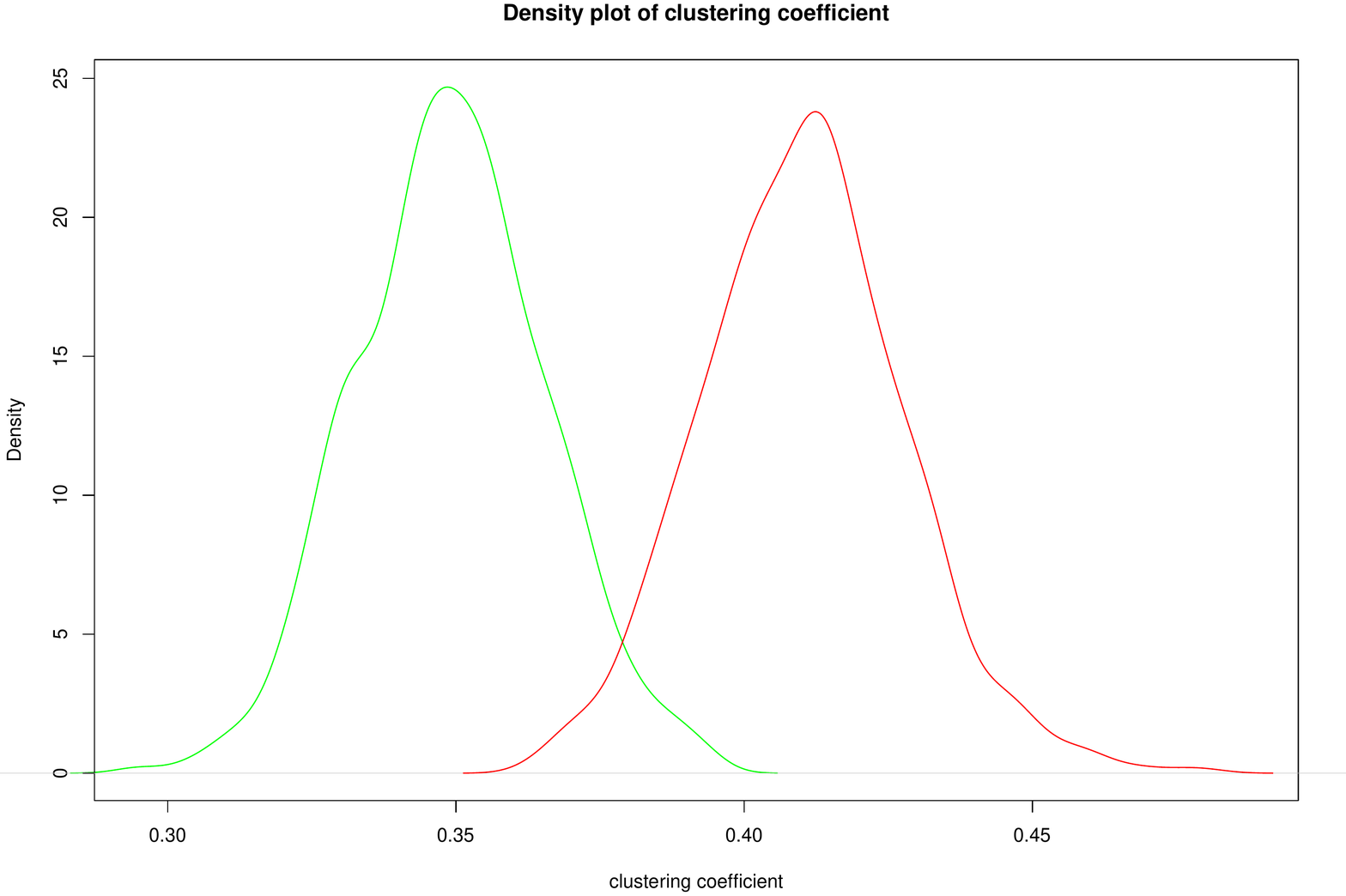}
		\caption{Density plot of the clustering coefficient in Simulation 1.}
	\end{subfigure}
     \caption{}
	    \label{sNC1}
\end{figure}

In Simulation 1, we take $N=200, \  K=3, \ \pi = (1/3, 1/3, 1/3), \ \Theta \equiv 1,$ and average degree $\lambda=15$. We use the "in-out-ratio" $r_1=p_1/q_1=20/3$ to generate 500 different networks, and $r_2=10$ to generate another 500 different networks. Then, we calculate the normalized clustering coefficient and the clustering coefficient for these 1000 networks to see whether we can identify these two types of networks through these two measures. The density curves for these two types of networks are plotted in  Figures \ref{sNC1} (a) and (b)). It can be seen that both measures can separate the two types of networks well.

\begin{figure}[H]
		\centering
	\begin{subfigure}[b]{0.45\linewidth}
		\includegraphics[width=\linewidth]{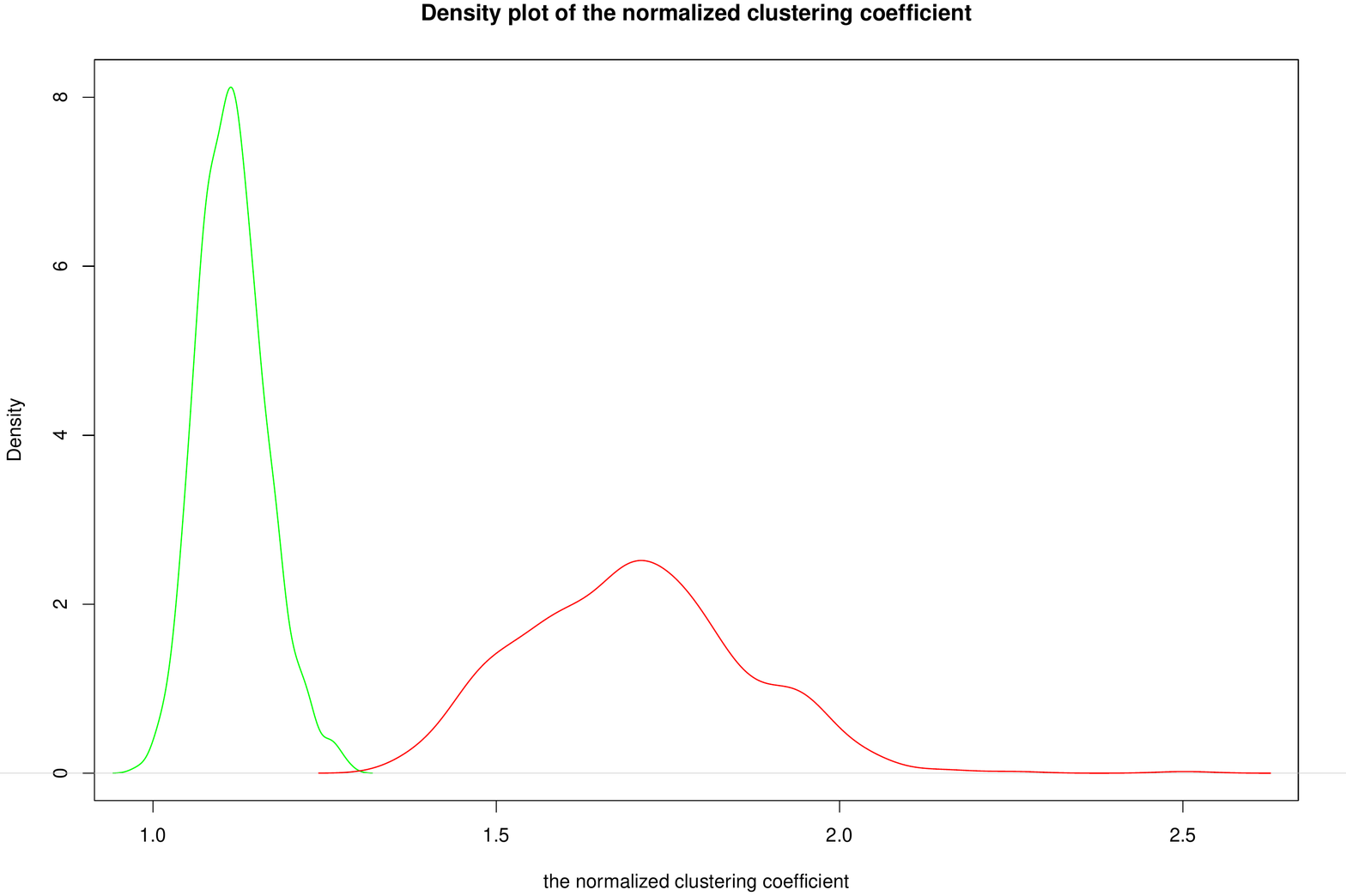}
		\caption{Density plot of the normalized clustering coefficient in Simulation 2.} 
	\end{subfigure}
	\begin{subfigure}[b]{0.45\linewidth}
		\includegraphics[width=\linewidth]{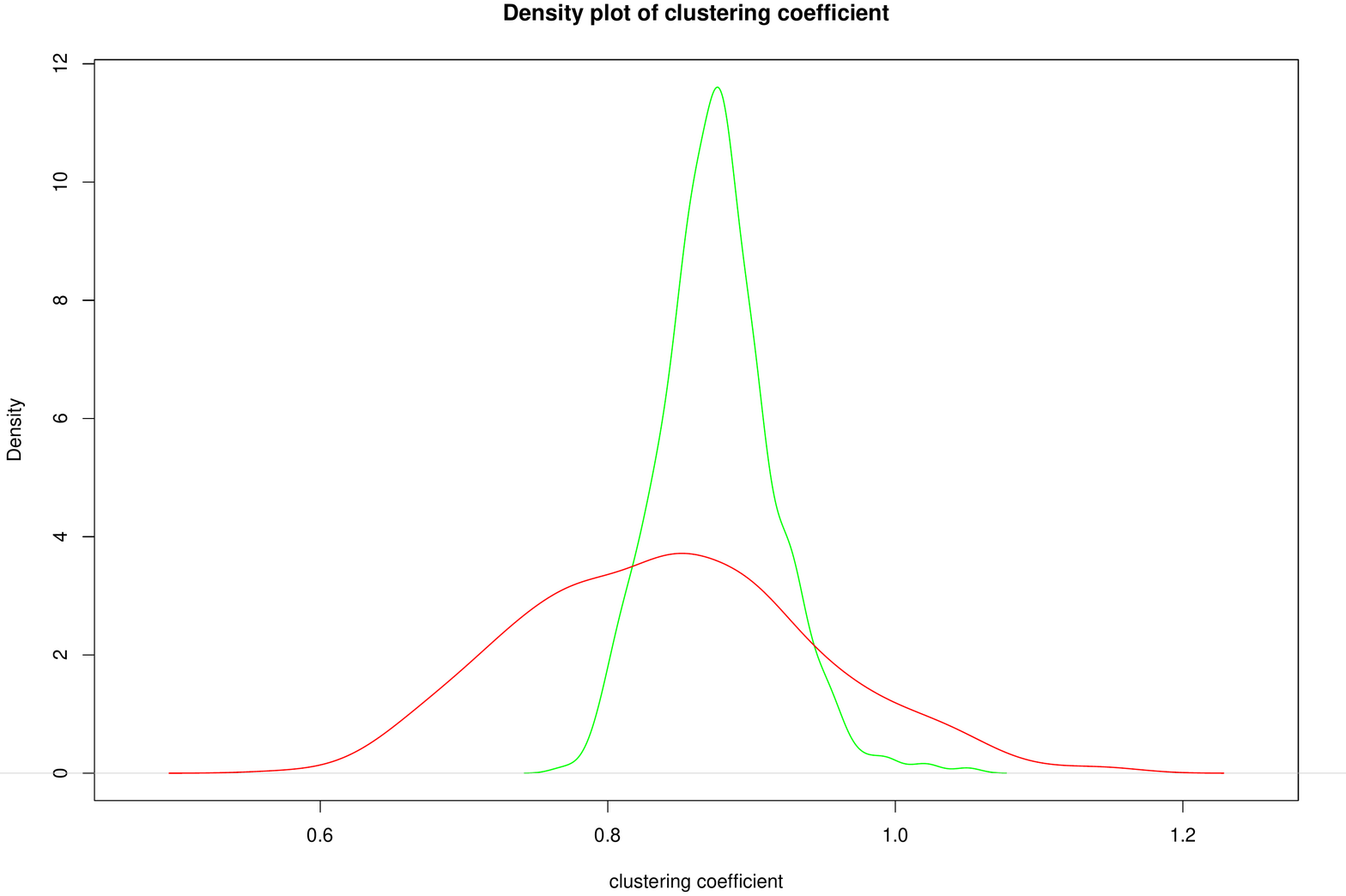}
	    \caption{Density plot of the clustering coefficient in Simulation 2.}
	\end{subfigure}
     \caption{}
	    \label{sNC2}
\end{figure}

Simulation 2 modifies the settings in Simulation 1. This time the $\Theta$ values are drawn independently with $P(\theta = 0.2) = 0.8$ and $P(\theta = 1) = 0.2$, corresponding to the DCBM with hubs. Meanwhile, 500 networks are sampled with the average degrees $\lambda_1$ and $\lambda_2$ drawn from uniform distributions $U(25,30)$ and $U(10,15)$, respectively. Other parameters are kept unchanged. Figures \ref{sNC2} 
(a) and (b) show that because of the effect of average degree and degree heterogeneity, the clustering coefficient cannot tell the difference between these two types of networks. However, the normalized clustering coefficient can still tell them apart. Furthermore, the last 500 networks all have larger normalized clustering coefficients than the first 500 networks, which is consistent with the ground truth.

\begin{figure}[H]
		\centering
	\begin{subfigure}[b]{0.45\linewidth}
		\includegraphics[width=\linewidth]{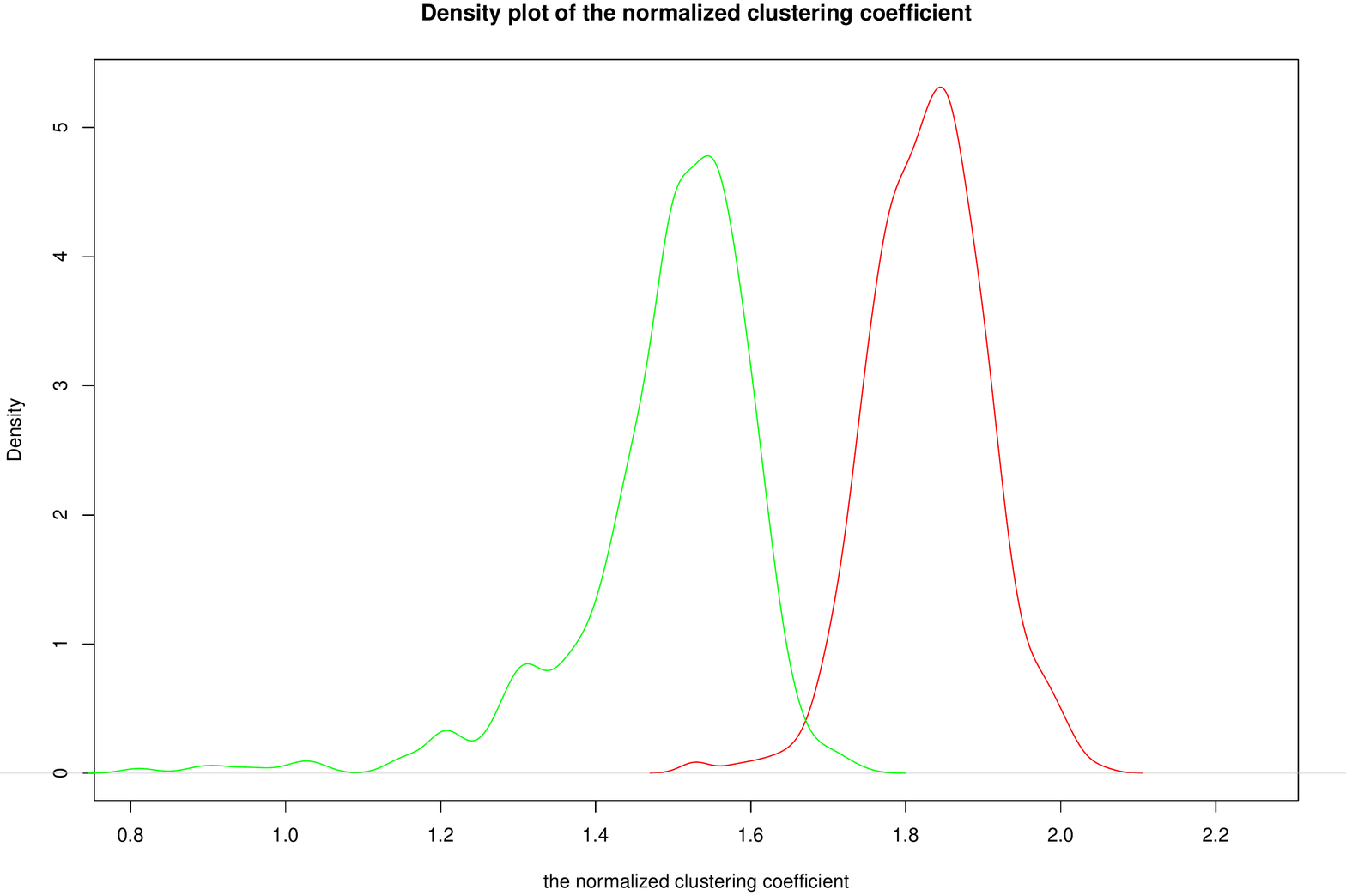}
		\caption{Density plot of the normalized clustering coefficient in Simulation 3.}
	\end{subfigure}
	\begin{subfigure}[b]{0.45\linewidth}
		\includegraphics[width=\linewidth]{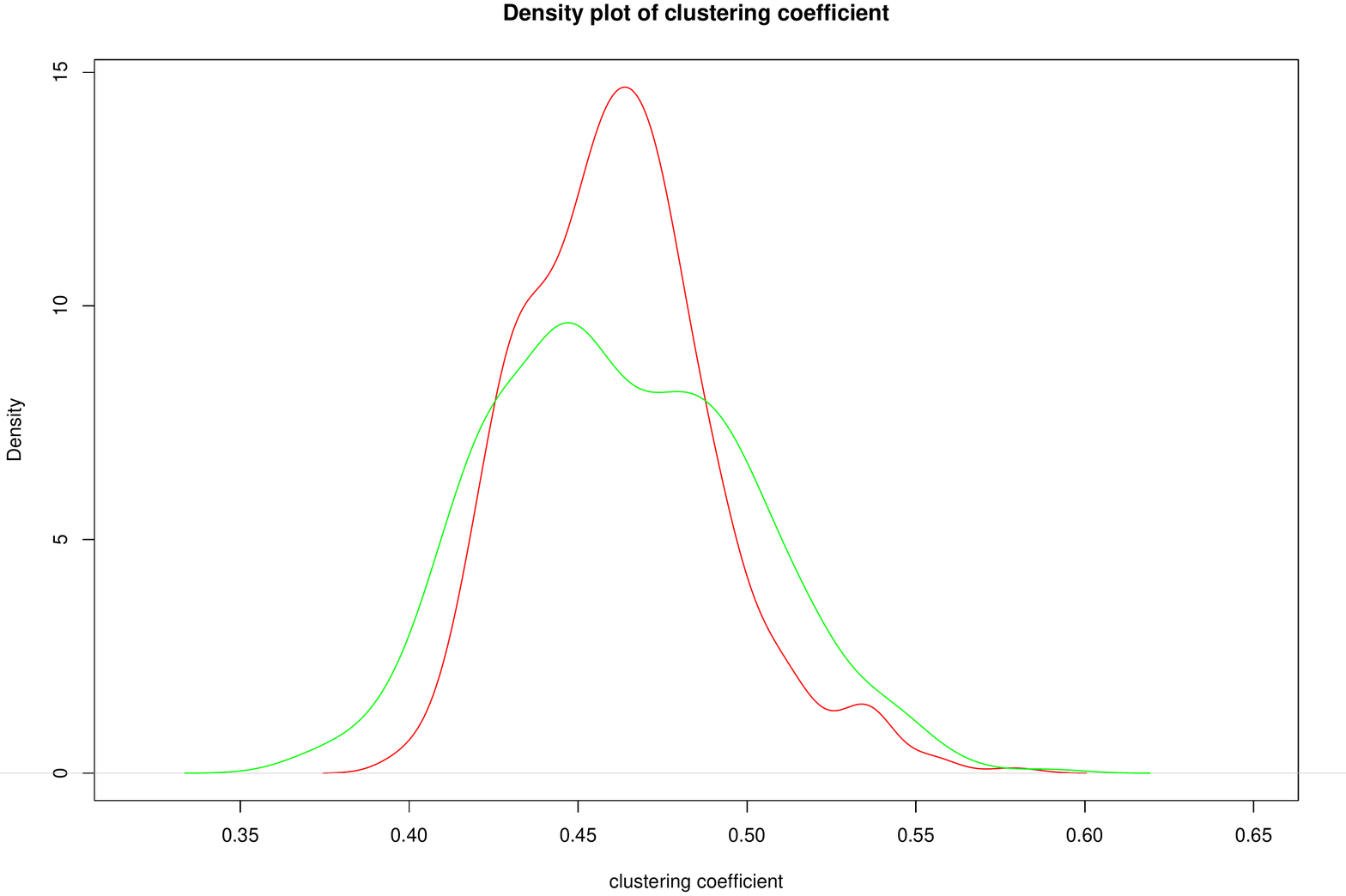}
		\caption{Density plot of the clustering coefficient in Simulation 3.}
    \end{subfigure}	
     \caption{}
	 \label{sNC3}
\end{figure}

In Simulation 3, different from Simulation 1, we make the distribution of $\Theta$ different for the two types of networks with $\Theta_1$ and $\Theta_2$ from two power law distributions with shape parameters $\alpha_1=4.2$ and $\alpha_2=6$ respectively. We keep other parameters the same as in Simulation 1. Figures \ref{sNC3}(a) and (b) show that the clustering coefficient could not remove the effect of the distribution of $\Theta$, while the normalized clustering coefficient could still catch the main difference between these two types of networks.   

From Simulations 1-3, it is clear that the average degree and the distribution of the degree of nodes do not affect the normalized clustering coefficient, while the clustering coefficient is badly affected. Furthermore, from Figures \ref{sNC1}-\ref{sNC3}, the normalized clustering coefficient is Gaussian with relatively small variance. All of these observations are consistent with the analysis in Section 2.

\subsection*{Real data on network clustering}

We download the data from

 \url{https://github.com/Kagandi/anomalous-vertices-detection/tree/master/data.}
  
  The network is constructed from the friendships between Twitter users in 2014 along with some labels of fake accounts. There are 5,384,160 nodes , among which 12,437 are labeled as fake accounts by Twitter. There are 16,011,443 edges in total. In \cite{rnet_real}, the authors propose an anomalous vertice detection algorithm based on link prediction. 

However, we first extract the one-step ego network of each node, which leads to 5,384,160 different networks in total. Then, we calculate the normalized clustering coefficient for each ego network and classify the nodes with this coefficient. We also compare the performance of the normalized clustering coefficient and the clustering coefficient. We focus on the nodes with degree larger than 20. In total, 148,830 users have more than 20 friends and 7,526 are labeled as fake. Among the 148,830 users, the ego networks of 52,031 do not have triangles or triplets and there are 4,473 fake users. Lastly, we focus on the remaining 96,799 accounts with 3,053 fake ids to compare the performance of the two statistics.

\section*{Appendix C} 
\subsection*{Network sampling algorithms}
 The network sampling algorithms we used in the simulations and empirical study are listed as follows:

Random node sampling (NS): Uniformly at random select a set of nodes. A sample is then a graph induced by the selected nodes.

Random edge sampling (ES): Uniformly at random select a set of edges. A sample is then a graph induced by the ends of these edges.

Random walk sampling (RWS): Uniformly at random pick a starting node and then simulate a random walk on the graph.

Random walk with flying back sampling (RWFS): Based on RWS, at every step with probability $p=0.15$ (the value commonly used in the literature \cite{leskovec2006}) we fly back to the starting node and re-start the random walk.

Random walk with jumping out sampling (RWJS): Based on RWS, at every step with probability $p=0.15$ (the value commonly used in the literature \cite{leskovec2006}) we randomly jump to any node in the graph and re-start the random walk.

Forest fire sampling (FF) (\cite{leskovec2006}): 

\textit{"We first choose node v uniformly at random. We then gen- erate a random number $x$ that is geometrically distributed with mean $p_f /(1-p_f )$. Node $v$ selects $x$ out-links incident to nodes that were not yet visited. Let $w_1,w_2,...,w_x$ denote the other ends of these selected links. We then apply this step recursively to each of $w_1,w_2,...,w_x$ until enough nodes have been burned. As the process continues, nodes cannot be visited a second time, preventing the construction from cycling. If the fire dies, then we restart it, i.e. select new node $v$ uniformly at random."} We take $p_f=0.7$ as suggested in \cite{leskovec2006}.

Snowball sampling (SS): We use function snowball\_sampling() in R package "netdep" (\url{https://cran.r-project.org/web/packages/netdep/index.html}).

\subsection*{Simulations on network sampling}
We perform simulations to compare different network sampling algorithms based on the normalized clustering coefficient. The better algorithms would be the ones that can obtain sub-networks whose normalized clustering coefficient is closer to that of  the original network. 

\begin{figure}[H]
	\centering
		\includegraphics[scale=.43]{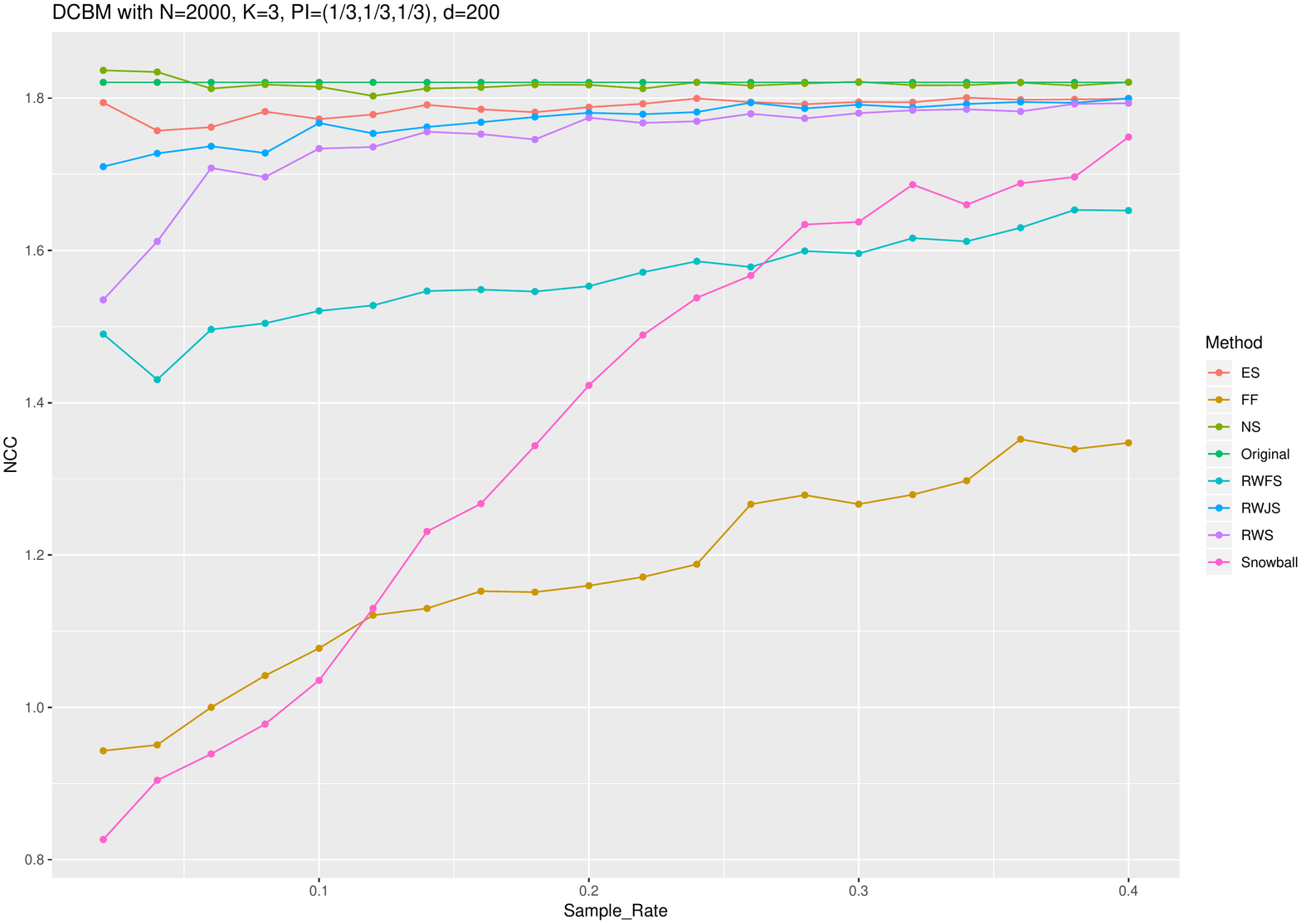}
		\caption{The normalized clustering coefficients of sub-networks and the original networks in Simulation 4. }\label{sampling_1}
\end{figure}

In Simulation 4, we generate a large network $G_1$ from the DCBM with parameters $N=2000, K=3, \ \pi = (1/3, 1/3, 1/3), \ r=p/q=10$, $\Theta$ generated from power law distribution and the average degree $d=200$. We let the ratio of sampling $f$ to vary from $2\%-40\%$. For each sampling algorithm and each sampling ratio, we sample 50 sub-networks and calculate their normalized clustering coefficient and then take the average for comparison with the original normalized clustering coefficient ($G_1's$). The result is plotted in Figure \ref{sampling_1}. It is clear from Figure \ref{sampling_1} that as $f$ increases, the normalized clustering coefficient of sub-networks becoming closer to the original one (light green line). NS and ES perform the best followed by RWJS and RWS. This is because $G_1$ is quite a dense network which reduces the risk of obtaining isolated nodes in the sub-network when NS or ES is applied. RWFS, FF and SS are unsatisfactory mainly because of the bias introduced by the initial nodes.   

\begin{figure}[H]
	\centering
	\includegraphics[scale=.43]{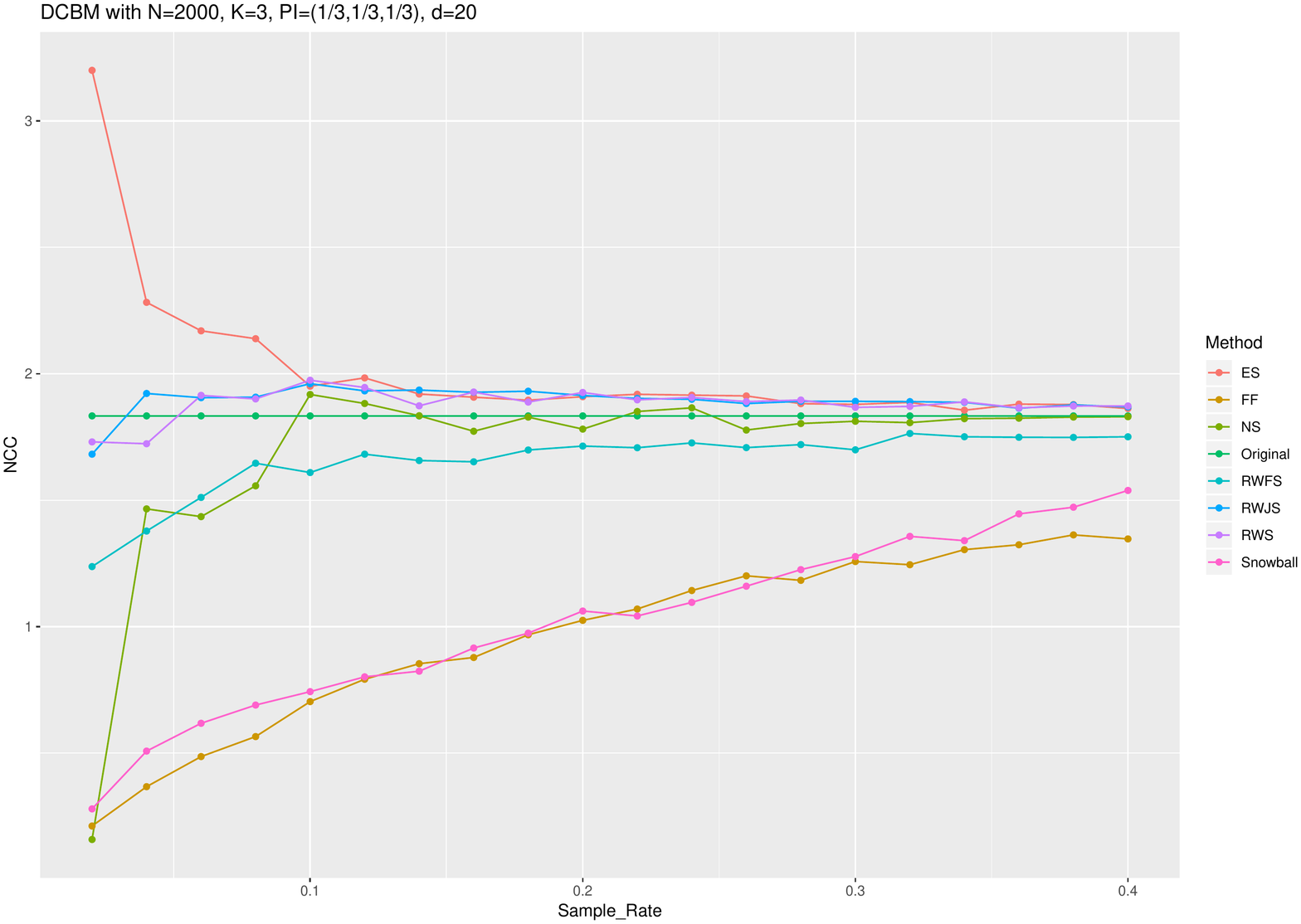}
	\caption{The normalized clustering coefficients of sub-networks and the original networks in Simulation 5. }\label{sampling_2}
\end{figure}

In Simulation 5, we reduce the average degree $d$ to 20 and obtain a new large network $G_2$. The result in Figure \ref{sampling_2} shows that all the algorithms perform worse than they do in Simulation 4, which indicates that sparsity increases the sampling difficulty. Moreover, the performance of NS and ES is greatly affected when the sampling ratio is less than  $10\%$.

\begin{figure}[H]
	\centering
	\includegraphics[scale=.43]{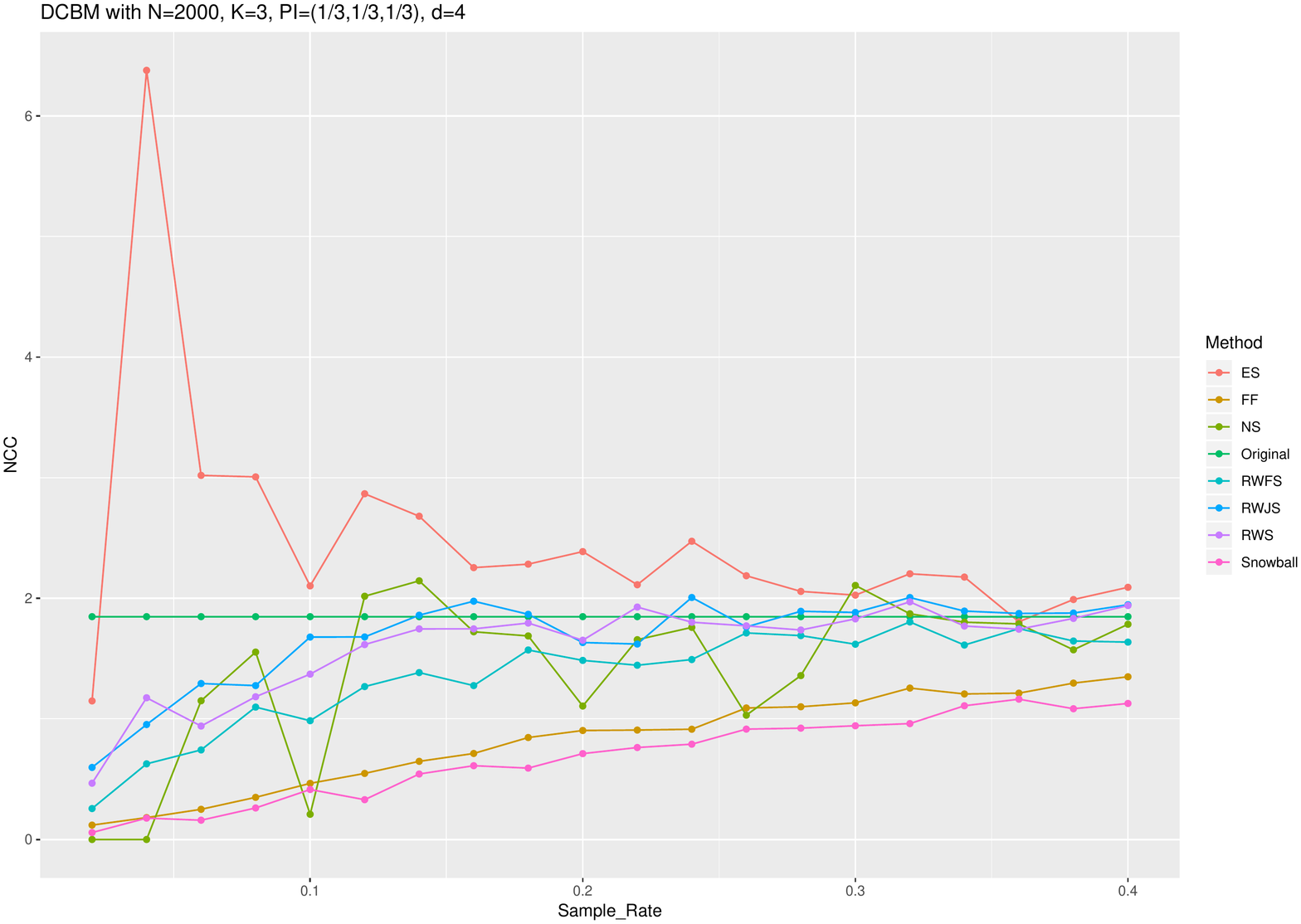}
	\caption{The normalized clustering coefficients of sub-networks and the original networks in Simulation 6. }\label{sampling_3}
\end{figure}

In Simulation 6, the performance of both NS and ES becomes worse when we reduce the average degree $d$ to $4$ to obtain a sparse network $G_3$ (see Figure \ref{sampling_3}). The standard deviations of NS and ES are quite large too. Most sub-networks drawn by NS and ES have many isolated nodes, which makes the performance unstable. However, RWJS and RWS continue to perform well in sparse networks. According to these simulations, RWJS and RWS can give a better trade-off between bias and variance than the other algorithms overall.

\subsection*{Real data on network sampling}
More details about the networks used in our empirical study on network sampling can be found on SNAP:

Condensed matter collaboration network (23133 nodes, 93497 edges):  \url{https://snap.stanford.edu/data/ca-CondMat.html}.

Enron email network (36692 nodes, 183831 edges): \url{http://snap.stanford.edu/data/email-Enron.html}.

Human protein-protein interaction network (21557 nodes, 342353 edges): \url{https://snap.stanford.edu/biodata/datasets/10000/10000-PP-Pathways.html}.

\section*{Appendix D} 
 The senate data we used in the dynamic network analysis are downloaded from \url{https://github.com/briatte/congress}. Edges are formed between two congress-persons using the weighted propensity (\cite{rsenate_generate}) to cosponsor with threshold $0.1;$ that is: let $Y_{ij(k)}=1$ if $i$ cosponsors $j$'s $k^{th}$ bill and 0 otherwise, and let $c_{j(k)}$ be the number of cosponsors of senator $j$'s $k^{th}$ bill. The relative weighted propensity to cosponsor of $i$ for $j$ is defined as
\begin{equation*}
WPC_{i,j}=\frac{\sum_{k=1}^{n_j}\frac{Y_{ij(k)}}{c_{j(k)}}}{\sum_{k=1}^{n_j}\frac{1}{c_{j(k)}}}
\end{equation*}
Lastly, we eliminate the edges with $WPC<0.1$ and set the edges with $WPC\geq 0.1$ to 1 to obtain an unweighted and undirected network for each year of senate.